\documentclass[a4paper, amsmath,amssymb,twocolumn,reprint, superscriptaddress,nofootinbib,
    ]{quantumarticle}
\pdfoutput=1

\usepackage[numbers,sort&compress]{natbib}
\usepackage{bbm}
\usepackage{graphicx}
\usepackage{csquotes}
\usepackage{xcolor}
\usepackage{mathtools}
\usepackage{physics}
\usepackage{booktabs}
\usepackage{siunitx}
\usepackage{float}
\usepackage{comment}
\usepackage{hyperref}
\usepackage{soul}

\graphicspath{{./figs}}

\newcommand{\fleft}{\mathopen{}\mathclose\bgroup\left}
\newcommand{\fright}{\aftergroup\egroup\right}
\newcommand{\kB}{k_\text{B}}
\newcommand{\ula}{\underline{a}}

\DeclareMathOperator{\diag}{diag}
\DeclareMathOperator{\artanh}{artanh}
\DeclareMathOperator{\mvec}{vec}

\newtheorem{theorem}{Theorem}[section]

\begin{document}

\title{Thermalization of open quantum systems with pseudomodes}

\author{R.~Kevin Kessing}
\email{kevin.kessing@uni-ulm.de}
\affiliation{Institut f{\"u}r Theoretische Physik, Universit{\"a}t Ulm, Albert-Einstein-Allee 11, 89081 Ulm, Germany}
\author{Thibaut Lacroix}
\email{thibaut.lacroix@uni-ulm.de}
\affiliation{Institut f{\"u}r Theoretische Physik, Universit{\"a}t Ulm, Albert-Einstein-Allee 11, 89081 Ulm, Germany}
\author{Susana F.~Huelga}
\email{susana.huelga@uni-ulm.de}
\affiliation{Institut f{\"u}r Theoretische Physik, Universit{\"a}t Ulm, Albert-Einstein-Allee 11, 89081 Ulm, Germany}
\affiliation{Center for Integrated Quantum Science and Technology (IQST), 89081 Ulm, Germany}
\author{Martin B.~Plenio}
\email{martin.plenio@uni-ulm.de}
\affiliation{Institut f{\"u}r Theoretische Physik, Universit{\"a}t Ulm, Albert-Einstein-Allee 11, 89081 Ulm, Germany}
\affiliation{Center for Integrated Quantum Science and Technology (IQST), 89081 Ulm, Germany}

\begin{abstract}
Pseudomode approaches allow for an exact and unapproximated description of a
quantum system interacting arbitrarily strongly with a bath.
In general, a system coupled to pseudomodes will not thermalize to the system's Gibbs state:
This is to be expected when the system--bath coupling is non-perturbative,
but conflicts with common thermodynamic intuition when the system--bath coupling is asymptotically weak.
We explore under which circumstances pseudomode models satisfy detailed balance
(and subsequently thermalize to the system Gibbs state)
and how specific choices of parameters can force ``weak'' detailed balance that is restricted to a limited frequency range.
A combination of Hermitian and non-Hermitian pseudomodes that yields a flat effective-temperature
profile is also considered.
The results and criteria established here are relevant for the construction of pseudomode models in contexts where thermodynamic consistency is required.
\end{abstract}

\maketitle

Since perfect isolation from all environmental influences is practically impossible, 
all quantum systems are, to varying extents, open quantum systems~\cite{BreuerPetruccione,Manzano2020AIPAdv,RivasHuelga}.
Unlike isolated systems, open quantum systems (if not subjected to further external influences)
typically feature one or more steady states that are reached in the long time limit:
The properties of these steady states are intrinsically linked
to the thermodynamic properties of the system and its environment~\cite{Spohn1977LMP,Frigerio1978CommMathPhys,SpohnLebowitz,Trushechkin2022AVS,DiMeglio2026JMP}.
When the system--environment coupling is nonperturbative,
the steady state of the system is not necessarily thermal~\cite{Iles-Smith2014PRA,Trushechkin2022AVS,Lambert2024PRR},
in the sense that the density matrix $\rho_S$ of the system does not necessarily converge towards the Gibbs state
of the system Hamiltonian $H_S$: $\lim_{t \to \infty} \rho_S \neq \frac{1}{Z} e^{-\beta H_S}$
with $Z = \Tr{e^{-\beta H_S}}$---a well-known fact that also holds true in classical statistical mechanics~\cite{Gibbs}.
On the other hand, as the system--environment coupling becomes vanishingly weak
and given an environment that is much larger than the system,
convergence to the Gibbs state of $H_S$ at the temperature of the environment
\emph{is} expected~\cite{Kosloff2013Entropy,Potts2024SciPost,Trushechkin2022AVS}.
\begin{figure}[t] 
\includegraphics[width=\linewidth]{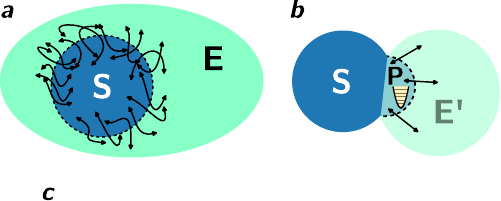}
\vspace{-4.1em}
\input{detbal_sketch}
\caption{%
The basic idea behind pseudomode models for open quantum systems.
\textit{\textbf{a}}:~A generic open quantum system (\textbf{S}) interacting strongly and non-trivially
with a structured environment (\textbf{E}).
Such complex environments are typically numerically inaccessible.
\textit{\textbf{b}}:~To capture the effect of \textbf{E} on \textbf{S} without having to treat \textbf{E} explicitly,
a pseudomode (\textbf{P}) is added to the system itself.
\textbf{P} is damped via a Lindblad dissipator~\eqref{eq:L_post}, representing
a simple environment \textbf{E'} whose effects on \textbf{P} are Markovian.
However, \textbf{P} interacts with \textbf{S} via the same coupling terms as the original \textbf{E} did---%
and, ideally, the combination of \textbf{P} and \textbf{E'} should reproduce the bath correlation functions of \textbf{E}.
\textit{\textbf{c}}:~A toy example of detailed balance. $E_j$ is the energy of the $j$-th system eigenstate.
The Gibbs state with populations $p_j \propto e^{-\beta E_j}$ is a steady state
if all transition rates $\Gamma(\omega)$ between eigenstates satisfy the detailed-balance condition
$\Gamma(\omega)/\Gamma(-\omega) = e^{\beta \hbar \omega}$.
}
\label{fig:pseudomode_illu}
\end{figure}%

Pseudomode approaches~\cite{Garraway1997PRA,Imamoglu1994PRA} (sketched in Fig.~\ref{fig:pseudomode_illu})
have become a prominent method for treating open quantum systems with strong system--bath 
coupling~\cite{Dorda2017NJP,Tamascelli2018PRL,Lemmer2018NJP,Mascherpa2020PRA,Weber2026NJP,Park2024PRB,Albarelli2025QST},
with wide-ranging applications reaching from strong light-matter coupling~\cite{Dalton2001PRA,Medina2021PRL,Sanchez2022Nanophot,Lednev2024PRL}
and energy transfer in organic photovoltaics and photosynthesis~\cite{Somoza2023CommPhys,Lorenzoni2024PRL,Lorenzoni2025SciAdv}
to trapped-ion simulators~\cite{Lemmer2018NJP,Sun2025NatComms,So2025NatComms} and
quantum heat engines~\cite{Albarelli2025QST,Weber2026NJP}.
However, contrary to intuitive expectations, reducing the system--pseudomode coupling to infinitesimal levels
does not, in general, restore detailed balance (see Fig.~\ref{fig:pseudomode_illu}\textbf{\textit{c}})---%
a central condition for thermodynamically consistent thermalization behavior~\cite{Agarwal1973ZPhys,Kossakowski1977CommMatPhys,Scandi2026PRX}.
Here, we examine under what conditions pseudomode approaches will or will not thermalize to the system's Gibbs state
$\frac{1}{Z} e^{-\beta H_S}$,
both when system--bath coupling is asymptotically weak and when it deviates from the weak-coupling limit.
We point out that even if the steady state of the system corresponds to a Gibbs state,
the relation between the temperature of the state and the temperature of the environment may be nontrivial,
a detail that has occasionally been noted in the past~\cite{Lambert2019NatComm,Lambert2024PRR,Lednev2024PRL},
but has not universally been taken into account~\cite{Albarelli2025QST}.
Our findings will provide a more rigorous, self-consistent framework
for quantum thermodynamics with pseudomodes in the weak- and strong-coupling regime
across several archetypal couplings.
As such, these results provide a clearer foundation for the use of pseudomodes in various situations 
in which the thermodynamic behavior and thermality or athermality
is central to a system, both in theoretical settings such as
quantum heat engines~\cite{alicki_quantum_1979,Hofer2017NJP,Weber2026NJP,deffner_quantum_2019},
mean-force Gibbs states~\cite{Miller2018,Trushechkin2022AVS},
and resource theory~\cite{chitambar_quantum_2019, lostaglio_introductory_2019},
as well as in experimental contexts such as (ultra-)strong light-matter coupling \cite{Kirton2015PRB,Du2021JCP,Satapathy2022AdvMat},
quantum reservoir engineering~\cite{Poyatos1996PRL,Plenio1999PRA,Plenio2002PRL,Kastoryano2011PRL,Lin2013Nature,Cormick2013NJP},
and analog quantum simulation of open quantum systems~\cite{Porras2008PRA,Lemmer2018NJP,MacDonell2021ChemSci,Olaya-Agudelo2025PRR}.

This paper is structured as follows.
In Section~\ref{sec:fundamentals}, we lay the foundation for our analysis by first recapitulating
the bath correlation functions (BCF) and corresponding effective dissipation rates of pseudomodes
and then introducing the notion of effective temperature based on detailed balance.
We also point out conditions for the uniqueness of the steady state,
which later enables us to identify under which circumstances detailed balance guarantees
thermalization to the system Gibbs state (or \enquote{return to equilibrium}~\cite{Trushechkin2022AVS}).
In Section~\ref{sec:weak_coupling}, we analyze the detailed balance and thermalization behavior of
pseudomode approaches under different types of system--bath coupling in the weak-coupling limit.
Here, we identify, across a variety of paradigmatic coupling models, the core issues that obstruct thermodynamic consistency
and then introduce the notion of \enquote{weak} detailed balance, which can be recovered by specific parameter choices.
In Section~\ref{sec:ho_example}, we present numerical results illustrating the
thermalizing or non-thermalizing behavior of the pseudomode approaches using an analytically solvable harmonic-oscillator example.
This validates the weak-coupling results of Section~\ref{sec:weak_coupling}
and complements them by extending the results to the strong-coupling regime.
Section~\ref{sec:nh} then shows how the \enquote{weak} detailed balance conditions can be strengthened
by extending the previously derived concepts to a multi-pseudomode model,
which thermalizes a system over a wider range of transition frequencies in the weak-coupling regime
while still demonstrating nonequilibrium steady states at stronger system--bath coupling.
We summarize our findings in Section~\ref{sec:conclusions}
and offer perspectives on how our analysis might apply to a range of other open-quantum-systems methods.


\section{Fundamentals} \label{sec:fundamentals}
In this first section, we reiterate the basics of pseudomodes
and then introduce the three categories of system--pseudomode couplings that we will treat in this paper.
We then derive one of the central pseudomode quantities---the bath correlation function---for these three models
and relate it to the notion of effective temperature, which will be central to the discussion of thermalization.
Finally, we point out conditions on the uniqueness of the steady state of a pseudomode model,
which allow us to formulate when a Gibbs state is not only \emph{a} valid steady state,
but also \emph{the} steady state of a system coupled to one or several pseudomodes.

\subsection{Pseudomodes}
A general master equation describing a system ($S$) under the influence
of a single pseudomode ($P$, illustrated in Fig.~\ref{fig:pseudomode_illu}) at any time~$t > 0$ is given by
\begin{equation}
\pdv{t} \rho_{SP} = -\frac{i}{\hbar}\comm{H_S + H_{SP}}{\rho_{SP}} + \mathcal{L}_P[\rho_{SP}],
\label{eq:rho_SP}
\end{equation}
where $H_S$ and $H_{SP}$ are the system and system--pseudomode interaction Hamiltonian,
respectively, and $\mathcal{L}_P$ is the pseudomode Lindbladian, which acts only on the pseudomode itself:
\begin{equation}\begin{multlined}
\mathcal{L}_P[\rho] = -\frac{i}{\hbar} \comm{H_P}{\rho}
    + N \gamma \left( a^\dag \rho a - \frac{1}{2} \acomm{a a^\dag}{\rho}\right)\\
    + (N + 1) \gamma \left( a \rho a^\dag - \frac{1}{2} \acomm{a^\dag a}{\rho}\right) .
\label{eq:L_post}
\end{multlined}\end{equation}
Here, $a$ (satisfying $\comm{a}{a^\dag} = \mathbbm{1}$) is the annihilation operator of the bosonic pseudomode
(which also acts as the jump operator of the dissipator),
$\beta = 1/(\kB T)$ is the inverse temperature of the bath acting on the pseudomode (the \emph{residual} environment),
$H_P = \hbar\omega_P a^\dag a$ is the pseudomode Hamiltonian and
\begin{equation}
    N = \frac{1}{e^{\beta\hbar\omega_P} - 1}
    \label{eq:Nbose}
\end{equation}
is the Bose--Einstein average occupation number of the pseudomode's energy $\hbar\omega_P$.
The generalization to several uncoupled pseudomodes is straightforward by adding indices $j$ to each
$a$, $\omega_P$, $N$ and $\gamma$.
For the remainder of the text, $N$ or $N_j$ will refer to a fixed pseudomode frequency $\omega_P$ or $\omega_{P,j}$,
respectively, as in~\eqref{eq:Nbose}, whereas $N(\abs{\omega})$ will be used when a variable frequency $\omega$ is meant.

Pseudomode models are governed by two defining features:
The nature of the system--pseudomode coupling Hamiltonian $H_{SP}$
and the bath correlation function (BCF) of the pseudomodes,
which we introduce in Sec.~\ref{sec:coupling_types} and~\ref{sec:bcf}, respectively.

\subsection{Three types of system--pseudomode coupling} \label{sec:coupling_types}
In this paper, we will consider the following three representative types of interactions $H_{SP}$:
First, an $x$-$x$-type model,
\begin{equation}
H_{SP} = \sum_{j} \hbar g_j \left(S_j + S_j^\dag\right) \otimes \left(a_j + a^\dag_j\right),
\label{eq:Ham_noRWA}
\end{equation}
where $S_j$ is some, typically non-Hermitian, system operator.
As $S_j$ is non-Hermitian, it will often induce energy transitions in the system.
Assume that $S_j$ has a relaxing (energy-decreasing) effect on the system and $S_j^\dag$
an exciting (energy-increasing) effect. 
Thus, we also investigate a version of~\eqref{eq:Ham_noRWA} with a specific rotating-wave approximation (RWA)
applied to it:
\begin{equation}
H_{SP} = \sum_j \hbar g_j \left(S_j \otimes a^\dag_j + S^\dag_j \otimes a_j\right)\, .
\label{eq:Ham_RWA}
\end{equation}
For example, in a quantum-optical setting, the non-RWA model~{\eqref{eq:Ham_noRWA}} may correspond to a Rabi model
and its RWA counterpart~{\eqref{eq:Ham_RWA}} to a Jaynes--Cummings model~{\cite{Xie2017JPA}}
if the pseudomode is taken to represent the influence of the electromagnetic cavity mode.

Furthermore, in many situations, the system--bath coupling is not derived from such $x$-$x$-type Hamiltonians
but is instead of the dephasing type
\begin{equation}
H_{SP} = \sum_j \hbar g_j Z_j \otimes \left(a_j + a^\dag_j\right),
\label{eq:Ham_zx}
\end{equation}
where $Z_j = Z_j^\dag$ is a system operator that is not generally exciting or relaxing.
For example, in spin systems, $Z_j$ may be proportional to $\sigma_z^{(j)}$,
or, in vibronic systems describing electron or excitation energy transfer,
$Z_j$ may be the projector $\dyad{e}$ onto the excited state.

The aforementioned requirements on $Z_j$ and $S_j$ can be expressed more precisely using
the standard system-eigenbasis representations of the system part of $H_{SP}$~\cite{RivasHuelga,Manzano2020AIPAdv},
\begin{equation}
S_j(\omega)
    = \sum_{\substack{\epsilon, \epsilon' \in \sigma(H_S)\\\epsilon' - \epsilon = \hbar\omega}}
            \Pi_\epsilon S_j \Pi_{\epsilon'},
    \label{eq:S_j_eigenbasis}
\end{equation}
where $\Pi_\epsilon = \sum_k \dyad{\epsilon, k}$ is the projector onto all system eigenstates of energy $\epsilon$
and $k$ is an index for degenerate energy levels. The spectrum $\sigma(H_S)$ is the set of all eigenvalues of $H_S$.
The sum is over all $\epsilon$ and $\epsilon'$ whose energy difference corresponds to a given
transition frequency $\omega$, and the set of all such $\omega$ under a given $H_S$ will be called the transition frequency
spectrum or Bohr (frequency) spectrum $\mathcal{B}$
\[\mathcal{B}
    = \left\{\omega \in \mathbb{R} \mid (\epsilon, \epsilon' \in \sigma(H_S))[\epsilon' - \epsilon = \hbar\omega] \right\}.\]
Using this framework, we can specify when $S_j$ may be called
a relaxing jump operator (or $S_j^\dag$ an exciting operator):
Any operator $A$ can be decomposed into its individual Bohr-spectrum components,
$A = \sum_{\omega \in \mathcal{B}} A(\omega)$, with $A(\omega)$ as defined in~\eqref{eq:S_j_eigenbasis}.
Since a typical operator does not mediate between all possible eigenstates,
$A(\omega)$ may be zero for many possible $\omega \in \mathcal{B}$.
Define the set of transition frequencies that \emph{are} mediated by the operator $A$ as $\mathcal{B}_A$:
\[\mathcal{B}_A \coloneqq \left\{\omega \in \mathcal{B} \mid A(\omega) \neq 0\right\}.\]
Then an operator $S_j$ is purely relaxing if it only mediates positive transition frequencies,
i.e., if $\mathcal{B}_{S_j} \subset \mathbb{R}_{> 0}$,
since, by convention, positive transition frequencies $\omega$ correspond to a decrease in energy.
Conversely, an operator $A$ is exciting if $\mathcal{B}_A \subset \mathbb{R}_{< 0}$
(note that $S_j$ being purely relaxing implies that ${S_j}^\dag$ is purely exciting).
If these conditions are satisfied, then~\eqref{eq:Ham_RWA} may reasonably be called a RWA, as, otherwise,
the contributions in~\eqref{eq:Ham_RWA} do not act in the typical \enquote{energy-conserving} manner of a RWA.
Furthermore, the diagonal operator $Z_j$ introduced in~\eqref{eq:Ham_zx} can be described
in this language as satisfying $\mathcal{B}_{Z_j} = \{0\}$.

\subsection{The pseudomode BCF and effective temperatures}\label{sec:bcf}
Much of our analysis is built on the concept of detailed balance,
illustrated in Fig.~\ref{fig:pseudomode_illu}\textbf{\textit{c}}.
In weakly coupled open quantum systems under \enquote{global} GKSL equations of the type
\begin{multline}
\pdv{\rho_S}{t} = -\frac{i}{\hbar} \comm{H_S + H_\text{LS}}{\rho_S}\\
    + \sum_{\omega \in \mathcal{B}_S} \Gamma(\omega) \left(S(\omega) \rho_S S^\dag(\omega) - \frac{1}{2} \acomm{S^\dag(\omega) S(\omega)}{\rho_S}\right)
\end{multline}
(where $H_\text{LS}$ is the Lamb--Stark Hamiltonian),
detailed balance is given by the condition
$\Gamma(\omega) = e^{\beta\hbar\omega} \Gamma(-\omega),$
which is always satisfied for such global GKSL dynamics via the Kubo--Martin--Schwinger (KMS) condition~\cite{SpohnLebowitz, RivasHuelga}.
If we invert this reasoning, we observe that the ratio $\Gamma(\omega)/\Gamma(-\omega)$
determines the \enquote{effective temperature}~\cite{Clerk2010RMP,Lambert2019NatComm,Lambert2024PRR},
\begin{equation}
T_\text{eff}(\omega) = \frac{\hbar\omega}{\kB \ln(\frac{\Gamma(\omega)}{\Gamma(-\omega)})},
\label{eq:T_eff}
\end{equation}
experienced by the system at some transition frequency $\omega$.
Accordingly, a system whose transition frequency spectrum $\mathcal{B}$ is contained in a narrow interval
around some $\omega'$ will approximately equilibrate to a thermal state
$\frac{1}{Z} \exp(-\frac{1}{\kB T_\text{eff}(\omega')} H_S)$
when the system--bath coupling is perturbatively small.
However, increasing the system--bath coupling will typically generate deviations from this equilibrating behavior,
as evidenced by the additional squeezing of the steady states shown in Sec.~\ref{sec:ho_example}.
Such behavior has also been studied in the past, e.g., in the context of mean-force Gibbs states~\cite{Miller2018,Trushechkin2022AVS}.

\begin{figure}[t!]
{\centering
\includegraphics[width=0.97\linewidth]{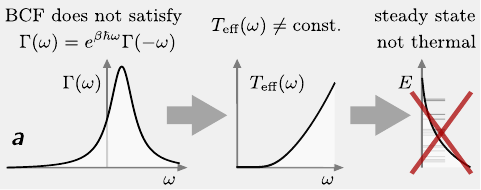}\\
\includegraphics{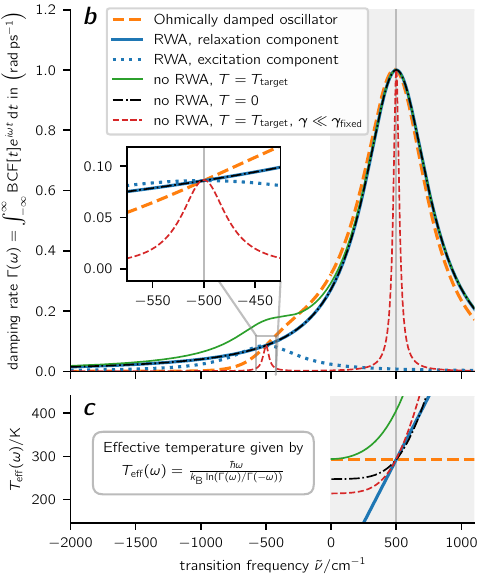}}
\caption{%
Damping rates and effective-temperature profiles for several single-pseudomode models.
\textit{\textbf{a}}:~A schematic explanation of the significance of the BCFs:
If the BCFs do not obey detailed balance, then the effective temperature~\eqref{eq:T_eff} is not constant
and the steady state is not a system Gibbs state $e^{-\beta H} / Z$, even in the weak-coupling limit.
\textit{\textbf{b}}:~%
The frequency-dependent damping rates $\Gamma(\omega)$---i.e., the Fourier-transformed BCFs---%
imposed upon the system by the single-pseudomode models.
In all cases, $\omega_P$ corresponds to $\tilde{\nu} = \SI{500}{\per\cm}$,
$g$ is chosen such that $\Gamma(\omega_P) = \SI{1}{\radian\per\ps}$
and $\gamma = \gamma_\text{fixed} \approx \SI{115}{\radian\per\ps}$ as given by~\eqref{eq:gamma_fixed}
except for the $\gamma \ll \gamma_\text{fixed}$ case where $\gamma = \SI{10}{\radian\per\ps}$.
The Ohmically damped harmonic oscillator resulting from $J_\text{ref}(\omega)$~\eqref{eq:Jref} (orange dashed line)
always satisfies detailed balance to $T_\text{target} = \SI{293}{\K}$ for all $\omega$.
The other BCFs only obey a \enquote{weak} detailed balanced at $\omega \approx \omega_P$.
The non-RWA/$T > 0$/large-$\gamma$ case (green line) is a counterexample
that demonstrates the artificial pumping induced by pseudomodes when parameters are not chosen carefully.
\textit{\textbf{c}}:~%
The effective temperatures $T_\text{eff}$ imposed upon the system by the various BCFs, as given by~\eqref{eq:T_eff}.
Only the Ohmically damped oscillator results in a constant $T_\text{eff} = T_\text{target}$,
with all other models never yielding the desired target temperature, or only at a single point.
}
\label{fig:basic_bcfs}
\end{figure}

The pseudomode bath correlation functions that are relevant for the coupling types introduced
in Sec.~\ref{sec:coupling_types} are,
for arbitrary times $t \in \mathbb{R}$ and residual-environment temperatures $T \geq 0$
(see App.~\ref{app:bcf_basics} for derivation):
\begin{align}\begin{split}
\expval{a(t) a^\dag(0)}
    &= e^{-i\omega_P t - \frac{\gamma}{2} \abs{t}} (N + 1), \label{eq:BCF_RWA}\\
\expval{a^\dag(t) a(0)}
    &= e^{i\omega_P t - \frac{\gamma}{2} \abs{t}} N ,
\end{split}\end{align}
where the parameters are as defined in~\eqref{eq:L_post}.
By extension, the correlation function of $x \coloneqq a + a^\dag$ with itself is
\begin{multline}
\expval{x(t) x(0)}
    = e^{-\frac{\gamma}{2} \abs{t}} \left(N e^{i\omega_P t} + (N + 1) e^{-i\omega_P t}\right)\\
    = e^{-\frac{\gamma}{2} \abs{t}} \left(\coth(\frac{\beta \hbar \omega_P}{2}) \cos(\omega_P t) - i\sin(\omega_P t)\right).
\label{eq:BCF_full}
\end{multline}
The correlation functions~\eqref{eq:BCF_RWA}~and~\eqref{eq:BCF_full} and their Fourier transforms
(the latter corresponding to the damping rates $\Gamma(\omega)$ in the weak-coupling limit)
represent the second defining feature of uncoupled pseudomodes.
A schematic summarizing the significance of the BCFs for thermalization is shown in Fig.~\ref{fig:basic_bcfs}\textit{\textbf{a}},
and the correlation functions that will play a role for our analysis are shown in Fig.~\ref{fig:basic_bcfs}\textit{\textbf{b}}.

A note on terminology: Throughout this paper, we will refer to the temperature contained within
the $N = 1/\left(e^{\beta \hbar \omega} - 1\right)$ of the pseudomode dissipator $\mathcal{L}_P$ of~\eqref{eq:L_post},
i.e., the temperature of the \emph{residual environment},
simply as $T$. The temperature experienced by the system itself will be referred to as the effective
temperature $T_\text{eff}$, and, where applicable, the \enquote{target temperature} will be called $T_\text{target}$,
which is set to ${T_\text{target} = \SI{293}{\K}}$ throughout.

\subsection{Uniqueness of the steady state}
The bulk of our analysis investigates whether the Gibbs state is a valid steady state of a given pseudomode model.
However, this alone does not guarantee convergence to the Gibbs state (i.e., thermalization),
as there may, in general, exist multiple steady states.
In App.~\ref{app:uniqueness}, we establish conditions under which the steady state is unique if it exists at all,
however, only as long as the residual temperature $T$ is non-zero (albeit potentially arbitrarily small).
In particular, Theorem~\ref{thm:uniqueness} establishes that, e.g.,
an $x$-$x$-type pseudomode model~\eqref{eq:Ham_noRWA} with an arbitrary system Hamiltonian $H_S$
either has no steady state at all or reaches a unique steady state if the only operators $A$ that satisfy
\[\forall\, j: \comm{A}{S_j + S_j^\dag} = \comm{A}{H_S} = 0\]
are those that are proportional to the identity: $A \propto \mathbbm{1}$.
For the RWA-type model~{\eqref{eq:Ham_RWA}}, the condition is similar: If
\[\forall\, j: \comm{A}{S_j} = \comm{A}{S_j^\dag} = \comm{A}{H_S} = 0\]
is true only for $A \propto \mathbbm{1}$, then the steady state is unique if it exists.
For $z$-$x$-type coupling~{\eqref{eq:Ham_zx}}, the Gibbs state is not guaranteed to be
(and often is not) the only steady state if $H_S$ and all $Z_j$ commute exactly,
but if at least one $Z_j$ only commutes approximately with $H_S$
such that $\comm{Z_j}{H_S} \neq 0$, then the same conditions apply as in the $x$-$x$-type case.

Furthermore, the uniqueness of the steady states of the example systems shown in Sec.~\ref{sec:ho_example} and~\ref{sec:nh}
is guaranteed (even for $T = 0$) by the properties of Gaussian states (see App.~\ref{app:uniqueness}).

If these uniqueness conditions are satisfied and a steady state exists,
then satisfying detailed balance is equivalent to thermalization to the Gibbs state.
Therefore, the following analysis will focus on detailed balance as the central condition for thermalization.


\section{The weak-coupling limit} \label{sec:weak_coupling}
We now consider the dynamics resulting from a pseudomode model
when the system--environment (i.e., system--pseudomode) coupling is made perturbatively weak.
First consider generic system--environment dynamics under a Hamiltonian
\[H_{SE} = H_S + H_E + \sum_j \hbar g_j A_{S,j} \otimes G_{E,j},\]
where $H_S$ and $H_E$ are the system and environment Hamiltonians that act on the
system ($\mathcal{H}_S$) and environment Hilbert space ($\mathcal{H}_E$), respectively, and the last set of terms
comprise the system--environment interactions with $A_{S,j} \colon \mathcal{H}_S \rightarrow \mathcal{H}_S$
and $G_{E,j} \colon \mathcal{H}_E \rightarrow \mathcal{H}_E$.
As described in Ref.~\cite{Tamascelli2018PRL}, if the bath correlation functions
$C^U_{jk}(t,s) = \expval{G_{E,j}(t) G_{E,k}(s)}$ of such a fully unitary model
are equal to pseudomode correlation functions from~\eqref{eq:BCF_RWA} and~\eqref{eq:BCF_full},
then the reduced \emph{system} dynamics under the pseudomode model exactly match those of the full unitary evolution.

In this section, we consider the limit of weak coupling between the system and the environment,
since many standard assumptions on canonical equilibrium thermodynamics are only known to hold in this limit~\cite{VanHove1957Physica,QuantumThermo2018,Rivas2020PRL,Talkner2020RMP,Trushechkin2022AVS}.
Additionally, this asymptotic weak-coupling/long-time regime corresponds to the van Hove limit,
in which one may expect a GKSL equation to become an increasingly accurate description of the system dynamics~\cite{Davies1974CommMath,Davies1976MathAnn,GKS1976JMP,Lindblad1976CommMathPhys,BreuerPetruccione}
(although we point out that there are formal difficulties in applying existing theory to unbounded system Hamiltonians
such as the harmonic oscillator considered in the following section).
Moreover, it is known that a GKSL equation that satisfies the detailed-balance conditions will lead to thermalization
to the system Gibbs state $\frac{1}{Z} e^{-\beta H_S}$ as long as the steady state is unique~\cite{RivasHuelga,Spohn1977LMP}.
Thus, the aim of this section is to investigate possible thermalization behavior of systems coupled
asymptotically weakly to pseudomodes by deriving effective GKSL equations describing
the perturbative action of the pseudomode on the system.
Note that these GKSL equations are used only to investigate the weak-coupling limit of the exact dynamics
described by the pseudomode approaches; the GKSL equations are in no way meant to reproduce the pseudomode dynamics in general.

To contrast the pseudomode behavior discussed in this paper from
that of \enquote{textbook} open quantum systems, we would like to highlight the following:
If an environment can be described in terms of a spectral density 
$J(\omega) = 2\pi \theta(\omega) \sum_k {g_k}^2 \delta(\omega - \omega_k)$
(where $\theta(\omega)$ is the Heaviside step function), then it effects a damping rate of
\begin{equation}
\Gamma(\omega) 
    = J(\omega) \left(N(\abs{\omega}) + 1\right) + J(-\omega) N(\abs{\omega})
\label{eq:Gamma_J}
\end{equation}
with $N(\abs{\omega}) = \left(e^{\beta \hbar \abs{\omega}} - 1\right)^{-1}$,
assuming the environment itself is in thermal equilibrium state at an inverse temperature $\beta$.
In such a situation, regardless of which $J(\omega)$ is chosen, the system Gibbs state $e^{-\beta H_S}/Z$
is always a steady state of a system that is weakly coupled to such a well-defined spectral density,
as, for any positive transition frequency $\omega_+$,
\begin{align*}
\Gamma(\omega_+) &= J(\omega_+) \left(N(\abs{\omega_+}) + 1\right) + \underbrace{J(-\omega_+) N(\abs{\omega_+})}_{=0},\\
\Gamma(-\omega_+) &= \underbrace{J(-\omega_+) \left(N(\abs{\omega_+}) + 1\right)}_{=0} + J(\omega_+) N(\abs{\omega_+}),
\end{align*}
where the second or first term, respectively, vanishes because of the $\theta(\omega)$ in $J(\omega)$.
Then the standard detailed-balance condition,
$\Gamma(-\omega) = e^{-\beta\hbar\omega}\Gamma(\omega)$, is always satisfied for all $\omega \in \mathbb{R}$.
Note that, crucially, this is the case because the occupation number $N(\abs{\omega})$ in~{\eqref{eq:Gamma_J}}
is a function of the transition frequency $\omega$, while in the pseudomode models,
$N$ as given in~\eqref{eq:Nbose} is a function of the fixed pseudomode frequency,
independent of the system transition frequency in question.
Furthermore, past research has investigated the differences between pseudomodes and a spectral density
that approximates the effect of a single pseudomode, which is the spectral density
induced by a unitarily damped harmonic oscillator~\cite{Garg1985JCP,Lemmer2018NJP}:
\begin{equation}
J_\text{ref}(\omega)
    = \frac{\gamma g^2 \theta(\omega)}{\frac{\gamma^2}{4} + \left(\omega_0 - \omega\right)^2}
        - \frac{\gamma g^2 \theta(\omega)}{\frac{\gamma^2}{4} + \left(\omega_0 + \omega\right)^2}
.
\label{eq:Jref}
\end{equation}
Pseudomode models can be shown to approximate this spectral density for $\gamma \ll \omega_0$.
The BCF resulting from $J_\text{ref}{(\omega)}$ is
shown as the dashed orange line in Figs.~{\ref{fig:basic_bcfs}}~and~{\ref{fig:nh_bcfs}}.
Given that the textbook open-system situation~\eqref{eq:Gamma_J} is guaranteed to thermalize
as a result of its KMS structure, one may wonder why pseudomodes,
which are known to be valid across all system--bath coupling regimes~\cite{Tamascelli2018PRL},
can fail to thermalize, as demonstrated in the following.
The resolution lies in the fact that the sum-of-Lorentzians BCF that a set of uncoupled pseudomodes generates
can only approximate a BCF of the type~\eqref{eq:Gamma_J}:
Indeed, as such a sum-of-Lorentzians BCF approaches a KMS-satisfying BCF, thermalization must be recovered.
However, achieving exact correspondence generally requires an arbitrarily high number of pseudomodes.
Any finite number of pseudomodes will introduce slight errors that may manifest in (non\nobreakdash-)thermalization
(although Sec.~\ref{sec:nh} shows a situation in which KMS structure is almost exactly restored over a finite region of $\omega$),
and, in particular, as the KMS condition depends on ratios between different frequencies,
even a correlation function that is well-matched in the time domain up to some finite time $t_\text{max}$
is not guaranteed to be well-matched in the frequency domains that are important for thermalization.

\subsection{With the RWA: Separate excitation and relaxation baths} \label{sec:rwa}
First, assume the RWA is applied, that is, the system--bath Hamiltonian is given by~\eqref{eq:Ham_RWA}.
In the weak-coupling limit, the system dynamics can be described by the following GKSL equation:
\begin{multline*}
\pdv{\rho_S}{t} = -\frac{i}{\hbar} \comm{H_S + H_\text{LS}}{\rho_S}\\
    + \sum_{j, \omega} \Gamma_j^{(r)}(\omega) \left(S_j(\omega) \rho_S S_j^\dag(\omega) - \frac{1}{2} \acomm{S_j^\dag(\omega) S_j(\omega)}{\rho_S}\right)\\
    + \sum_{j, \omega} \Gamma_j^{(e)}(\omega) \left(S_j^\dag(\omega) \rho_S S_j(\omega) - \frac{1}{2} \acomm{S_j(\omega) S_j^\dag(\omega)}{\rho_S}\right),
\end{multline*}
where the $\omega$ sums are over $\mathcal{B}_{S_j}$ and $\mathcal{B}_{S_j^\dag}$, respectively, and
the damping rates, assuming uncoupled baths, are
\begin{align*}
\Gamma_j^{(r)}(\omega) &= {g_j}^2 \int_{-\infty}^\infty \expval{a_j(t) a^\dag_j(0)} e^{i\omega t} \dd{t},\\
\Gamma_j^{(e)}(\omega) &= {g_j}^2 \int_{-\infty}^\infty \expval{a^\dag_j(t) a_j(0)} e^{i\omega t} \dd{t}.
\end{align*}
The superscripts $r$ and $e$ here represent relaxation and excitation, respectively,
reflecting the relaxing/exciting nature of $S_j$ and $S_j^\dag$ mentioned above.
If the system--bath interaction can be described by pseudomodes,
these damping rates are proportional to the two-sided Fourier transforms
of the correlation functions~\eqref{eq:BCF_RWA}. Specifically, if each index $j$ corresponds to a single pseudomode, then
\begin{align}\begin{split}
\Gamma_j^{(r)}(\omega) 
    &= {g_j}^2 \left(N_j + 1\right) \mathfrak{L}_j(\omega)\\
\Gamma_j^{(e)}(\omega) 
    &= {g_j}^2 N_j \mathfrak{L}_j(-\omega),
\label{eq:BCF_pseudomode_wRWA}
\end{split}\end{align}
(see~Fig.~\ref{fig:basic_bcfs}\textit{\textbf{b}}) where
\[\mathfrak{L}_j(\omega) 
    \coloneqq \frac{\gamma_j}{\frac{{\gamma_j}^2}{4} + \left(\omega - \omega_{P,j}\right)^2}\]
and $N_j$ is as given in~\eqref{eq:Nbose} for a pseudomode frequency $\omega_{P,j}$
(note that, crucially, $N_j$ depends on $\omega_{P,j}$ and not on the transition frequency $\omega$).
Therefore, for weak coupling, the pseudomode system with the RWA is equivalent to
an identical same system coupled to independent heat baths for each relaxing $S_j$ and exciting $S_j^\dag$.
The coupling to these relaxation and excitation baths is given by Lorentzian BCFs centered at some positive and negative frequency, respectively.

Assume for now that only a single pseudomode couples to the system, i.e., that there is only a single index $j$.
Given that system excitations should be mostly induced by $S_j^\dag$ and relaxations by $S_j$,
detailed balance in this situation depends primarily on the following ratio:
\begin{equation}
\frac{\Gamma_j^{(r)}(\omega)}{\Gamma_j^{(e)}(-\omega)} = \frac{N_j + 1}{N_j} = e^{\beta \hbar\omega_{P,j}}.
\label{eq:detbal_RWA}
\end{equation}
Thus, these correlation functions fulfill a \enquote{weak} detailed-balance condition,
meaning that detailed balance to the target temperature is only approximately valid
for transition frequencies $\omega$ close to $\omega_{P,j}$,
that is, if $\mathcal{B}_{S_j} = \{\omega_{P,j}\}$ or if all elements of $\mathcal{B}_{S_j}$ are close to $\omega_{P,j}$.
More generally, if all elements in $\mathcal{B}_{S_j}$ lie close to some frequency $\omega_0$,
the RWA pseudomode can be tuned to accommodate this frequency $\omega_0$:
Since $\Gamma_j^{(r)}(\omega) = e^{\beta \hbar\omega_{P,j}} \Gamma_j^{(e)}(-\omega)$ for an arbitrary $\omega$,
we can obtain detailed balance around $\omega_0$ at an inverse target temperature $\beta'$
by using a specifically chosen pseudomode bath temperature $\beta$:
\begin{equation}
\beta = \beta' \frac{\omega_0}{\omega_{P,j}}\\
     \implies \Gamma_j^{(r)}(\omega) \approx e^{\beta' \hbar \omega} \Gamma_j^{(e)}(-\omega)
\label{eq:Btemp}
\end{equation}
as long as $\omega \approx \omega_0$.

If several pseudomodes are coupled to the same jump operator, corresponding to a system--environment Hamiltonian of the type
\[S \otimes \sum_j \hbar g_j a^\dag_j + S^\dag \otimes \sum_j \hbar g_j a_j,\]
then the combined BCFs $\sum_j \Gamma_j^{(r)}$ and $\sum_j \Gamma_j^{(e)}$ are experienced by the system
and the detailed-balance condition becomes
\[\frac{\sum_j \Gamma_j^{(r)}(\omega)}{\sum_j \Gamma_j^{(e)}(-\omega)}
    = \frac{\sum_j {g_j}^2 \left(N_j + 1\right) \mathfrak{L}_j(\omega)}{\sum_j {g_j}^2 N_j \mathfrak{L}_j(\omega)}
    \overset{!}{=} e^{\beta\hbar\omega}.\]
If all $\omega_{P,j} \approx \omega$, then~\eqref{eq:detbal_RWA} may still approximately hold true,
but in general, a collection of pseudomodes will not cause the system to thermalize to the residual
environment's temperature without specific fine-tuning.

In summary, if the RWA is applied to the system--pseudomode coupling,
then weak detailed balance is satisfied for transition frequencies close to the pseudomode frequency $\omega_0$,
and weakly-coupled systems whose transition frequency is equal to $\omega_0$ will thermalize to the pseudomode temperature.
The frequency/temperature combination can be tuned using~\eqref{eq:Btemp}.

\subsection{Without the RWA: Effective temperature via damping parameters}
On the other hand, if the RWA is not applied, that is, the system--bath Hamiltonian is given by~\eqref{eq:Ham_noRWA},
then the corresponding GKSL equation would be (defining ${X_j \coloneqq S_j + S_j^\dag = X_j^\dag}$)
\begin{multline}
\pdv{\rho_S}{t} = -\frac{i}{\hbar} \comm{H_S + H_\text{LS}}{\rho_S}\\
    + \sum_{j, \omega} \Gamma_j(\omega) \Bigg[X_j(\omega) \rho_S X_j(\omega)
        - \frac{1}{2} \acomm{X_j(\omega) X_j(\omega)}{\rho_S}\Bigg]
\label{eq:GKSL_noRWA}
\end{multline}
where the $\omega$-sum is now over all $\omega \in \mathcal{B}_{X_j}$.
Here, \eqref{eq:BCF_full}~is the relevant correlation function and its Fourier transform
is found to be (see~Fig.~\ref{fig:basic_bcfs}\textit{\textbf{b}})
\begin{equation}
\Gamma_j(\omega) = {g_j}^2 \left[ \left(N_j + 1\right) \mathfrak{L}_j(\omega) + N_j\, \mathfrak{L}_j(-\omega)\right].
\label{eq:BCF_pseudomode_woRWA}
\end{equation}
Clearly, this does not fulfill the same weak detailed-balance condition, even for a single pseudomode, since
$\Gamma_j(-\omega)
    = \left(e^{-\beta\hbar\omega_{P,j}} + \epsilon \right) \Gamma_j(\omega)$,
where $\epsilon$ is an \enquote{error term}:
\[\epsilon = \frac{2 \sinh(\beta\hbar\omega_{P,j}) \left(\frac{{\gamma_j}^2}{4} + \left(\omega - \omega_{P,j}\right)^2\right)}{%
    \left(\frac{{\gamma_j}^2}{4} + \left(\omega + \omega_{P,j}\right)^2 \right) e^{\beta\hbar\omega_{P,j}}
    + \frac{{\gamma_j}^2}{4}
    + \left(\omega - \omega_{P,j}\right)^2
    }.\]
Strict detailed balance is never satisfied, but weak detailed balance for $\omega \approx \omega_{P,j}$ is satisfied if $\epsilon \to 0$.
This is asymptotically true in the rather trivial high-temperature limit $\beta \hbar \omega_{P,j} \to 0$,
as well as in the weakly-damped-pseudomode limit, 
\begin{equation}
\gamma_j \to 0 \qq{with} \omega \approx \omega_{P,j}.
\label{eq:small_gamma}
\end{equation}
The latter case can be interpreted as a dissipation rate $\Gamma(\omega)$ that consists of two Lorentzians,
one centered at $\omega_{P,j}$ and one at $-\omega_{P,j}$, whose width $\gamma$ is sufficiently narrow that they barely overlap.
Then the value of $\Gamma(+\omega)$ is dominated by the positive-frequency Lorentzian and vice versa,
and the ratio $\Gamma(\omega_{P,j})/\Gamma(-\omega_{P,j})$ is approximately $\frac{N_j + 1}{N_j} = e^{\beta\hbar\omega_{P,j}}$,
as is shown by~\eqref{eq:BCF_pseudomode_woRWA}.

However, an alternative method to enforce a weak detailed-balance condition even for non-zero $\gamma_j$
can be found by setting $T = 0$ and fixing $\gamma_j$ to a specific value.
Note that for $T = 0$:
\begin{equation}
\Gamma_j(\omega)
    = {g_j}^2 \mathfrak{L}_j(\omega)
    = {g_j}^2 \frac{\gamma_j}{\frac{{\gamma_j}^2}{4} + \left(\omega - \omega_{P,j}\right)^2}.
\label{eq:BCF_pseudomode_woRWA_T0}
\end{equation}
We can then enforce the weak detailed-balance condition to a specific $\beta$ for $\omega \approx \omega_{P,j}$ by requiring
${\Gamma_j(-\omega_{P,j}) \overset{!}{=} e^{-\beta\hbar\omega_{P,j}} \Gamma_j(\omega_{P,j})}$.
This fixes $\gamma_j$ to the value
\begin{equation}
\gamma_j = \frac{4 \omega_{P,j}}{\sqrt{e^{\beta\hbar\omega_{P,j}} - 1}} \eqqcolon \gamma_\text{fixed}.
\label{eq:gamma_fixed}
\end{equation}

Furthermore, coupling multiple pseudomodes to the same system operator, as in
\[\left(S + S^\dag\right) \otimes \sum_j \hbar g_j \left(a_j + a^\dag_j\right),\]
once again results in the BCFs given in~\eqref{eq:BCF_pseudomode_woRWA} being added together, so that the 
condition for detailed balance becomes
\[\frac{\sum_j {g_j}^2 \left[\left(N_j + 1\right) \mathfrak{L}_j(\omega) + N_j \mathfrak{L}_j(-\omega)\right]}{%
    \sum_j {g_j}^2 \left[\left(N_j + 1\right) \mathfrak{L}_j(-\omega) + N_j \mathfrak{L}_j(\omega)\right]}
    \overset{!}{=} e^{\beta\hbar\omega}.\]
This condition is, naturally, not generally fulfilled, but the parameters can be chosen in a specific
way so as to approximately satisfy detailed balance over a region of $\omega$.
We also note that, e.g., adding a single negative-frequency mode at $-\omega_{P,j}$ with a
modified coupling strength $\alpha g_j$ does not resolve the problems created by a single mode at $+\omega_{P,j}$
(see App.~\ref{app:negative_mode} for details).

Past analyses have highlighted that---if one considers pseudomodes to be derived from,
or approximations of, unitarily damped harmonic oscillators---typical pseudomode methods effectively drop Matsubara terms
that would otherwise appear~\cite{Lemmer2018NJP}, and that reinstating these Matsubara terms could
also remedy the broken detailed-balance conditions~\cite{Lambert2019NatComm,Lambert2024PRR}.
However, technical difficulties arise when dealing with these Matsubara terms, and, moreover,
derivations such as~\cite{Tamascelli2018PRL} make no assumptions about the unitary system
from which a pseudomode model is derived.

To summarize the $x$-$x$-type coupling: Achieving detailed balance in this scenario is more difficult
than in the RWA case of Sec.~\ref{sec:rwa} due to the \enquote{artificial pumping} effect~\cite{Lednev2024PRL}
attributable to the long tails of the Lorentzians comprising the BCF.
Both the frequency and the damping rate need to be fine-tuned even for weak detailed balance;
if the system Bohr spectrum is localized around some $\omega_0$, then a weakly $x$-$x$-coupled pseudomode
will cause the system to approximately thermalize to its residual temperature if the damping rate satisfies~\eqref{eq:gamma_fixed}.

\subsection{\texorpdfstring{$z$-$x$}{z-x} system--bath coupling}
Assume now that $H_{SP}$ is given by the $z$-$x$-type coupling of~{\eqref{eq:Ham_zx}}
with the system part of $H_{SP}$ described by some $Z_j$.
In the pure-dephasing case, only zero-frequency transitions are mediated by $Z_j$,
i.e., $Z_j(\omega) = 0$ for any $\omega \neq 0$ (equivalently, $\comm{H_S}{Z_j} = 0$ and $\mathcal{B}_{Z_j} = \{0\}$),
and then the Gibbs state---along with any other fully incoherent mixture of system eigenstates---%
is always a valid steady state of the GKSL equation.

However, in practice, the exact commutativity of $Z_j$ and $H_S$ may be broken by, e.g.,
weak inter-site coupling terms in $H_S$.
In such a situation, the weak-coupling limit GKSL equation would be identical to~\eqref{eq:GKSL_noRWA} after replacing
$S_j(\omega) + S_j^\dag(\omega) \mapsto Z_j(\omega)$, and
with the dissipation rates $\Gamma_j(\omega)$ given by~\eqref{eq:BCF_pseudomode_woRWA} as well.
Then the same arguments that apply to the non-RWA case also apply to the $z$-$x$ case
and detailed balance will also not be satisfied except under the aforementioned special circumstances---%
however, a restriction to low-frequency transitions may still apply:
As discussed in Sec.~\ref{sec:coupling_types}, the $Z_j$ operator often still does not mediate large energy transitions,
that is, all Bohr frequencies $\omega$ of $\mathcal{B}_{Z_j}$ have a small absolute value $\abs{\omega}$
and only the low-transition-frequency component of $\Gamma(\omega)$ is \enquote{visible} to $Z_j(\omega)$.
If there is some $\omega_c$ such that $\abs{\omega} < \omega_c$ for all $\omega \in \mathcal{B}_{Z_j}$,
then any potentially athermal nature of the bath correlation function for $\abs{\omega} > \omega_c$ will not
be experienced by the system and $\Gamma(\omega)$ only needs to be thermal in the range $(-\omega_c, \omega_c)$
to effectively provide detailed balance to the system.


\begin{figure*}[t]
\includegraphics{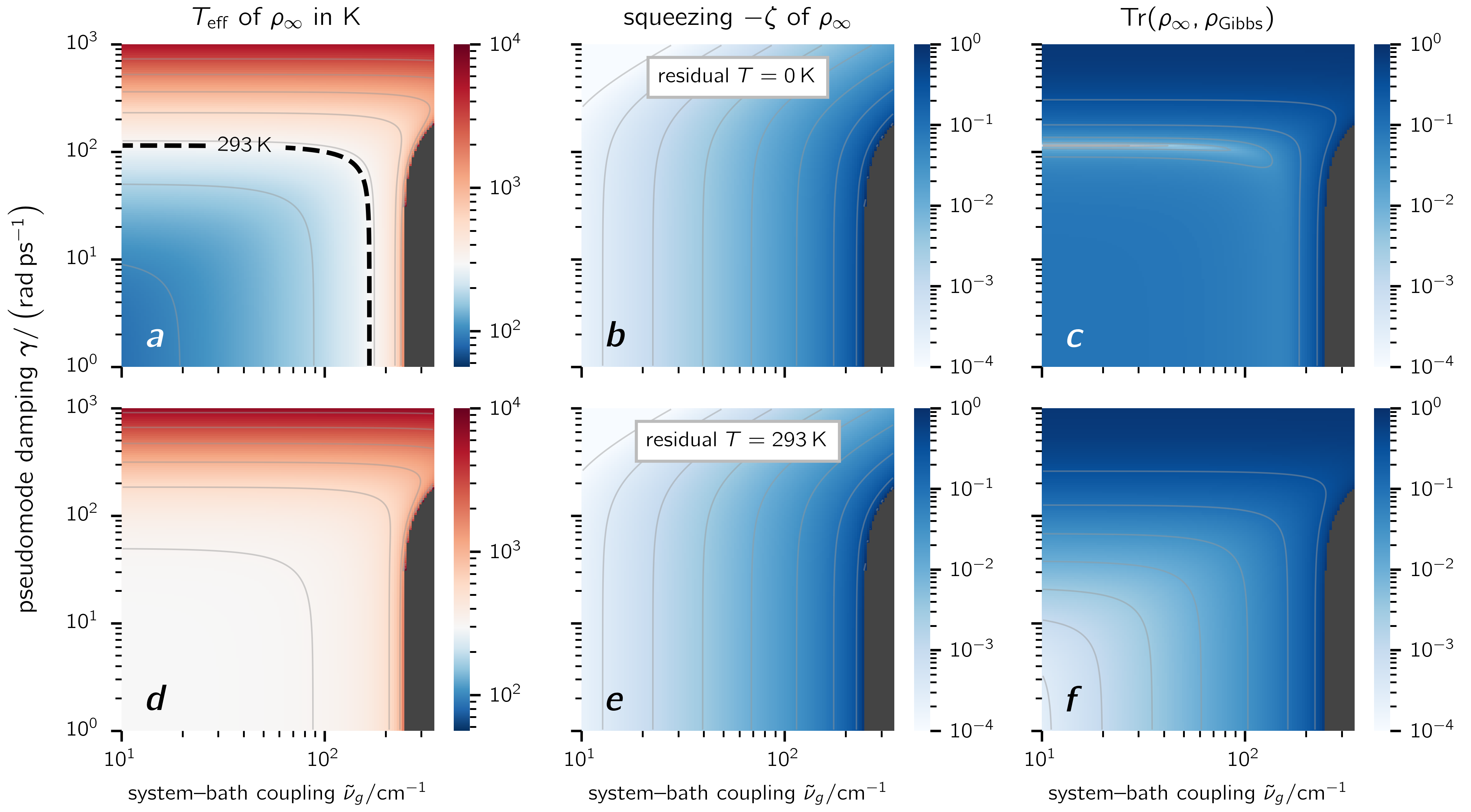}
\caption{Steady-state properties of a harmonic oscillator ($\tilde{\nu}_0 = \SI{500}{\per\cm} \iff \omega_0 \approx \SI{94}{\radian\per\ps}$)
that is $x$-$x$-coupled~{\eqref{eq:Ham_noRWA}} to a single resonant ($\omega_0 = \omega_P$) pseudomode
at a residual-environment temperature of $T = \SI{0}{\K}$ (top row, \textbf{\textit{a}}--\textbf{\textit{c}})
or $T = T_\text{target} = \SI{293}{\K}$ such that $\hbar\omega_0 \approx 2.5 \kB T_\text{target}$
(bottom row, \textbf{\textit{d}}--\textbf{\textit{f}}).
The coupling $g$ is given in terms of its wavenumber $\tilde{\nu}_g = g / (2\pi c_\text{vac})$.
The prominent area around $\gamma \approx \SI{115}{\radian\per\ps}$ for small $g$ in the $T = 0$ cases
corresponds to $\gamma_\text{fixed}$~\eqref{eq:gamma_fixed}
and is also shown as the black dash-dotted line in Fig.~{\ref{fig:basic_bcfs}},
while the small-$\gamma$ regions of the bottom row ($T = T_\text{target}$)
corresponds to the black dashed line in Fig.~{\ref{fig:basic_bcfs}}.
The gray areas mark the divergent regimes in which the system--bath coupling $g$ is larger than
the critical value $\frac{1}{2} \sqrt{{\omega_0}^2 + \frac{\gamma^2}{4}}$~\eqref{eq:g_crit}.
\emph{Left-hand column:}
The effective temperature $T_\text{eff} = 1 / (\kB \tilde{\beta})$ of the steady state, as given in~\eqref{eq:beta_tilde}.
Ideally, $T_\text{eff}$ should equal $T_\text{target} = \SI{293}{\K}$,
which corresonds to white on the logarithmic color scale.
The contour lines are located at $10^{n/4}\,\si{\K}$ for integer $n$,
plus a dashed contour line at $T_\text{target}$.
Note that~\eqref{eq:T_eff} only predicts $T_\text{eff}$ for small $g$ (left-hand end of the diagrams)
and that deviations from the predicted $T_\text{eff}$ are non-secular/non-perturbative effects.
\emph{Center column:}
The negative of the squeezing parameter $\zeta$~\eqref{eq:squeezing}.
The steady state is only truly thermal when $\zeta \to 0$. Contour lines at $10^{-n/2}$ for integer $n$.
\emph{Right-hand column:}
The trace distance~\cite{NielsenChuang} between the steady state and the Gibbs state at $T_\text{target} = \SI{293}{\K}$.
When this value is small, the system approximately thermalizes to the target temperature.
Contour lines at $10^{-n/2}$ for integer $n$.
}
\label{fig:combined}
\end{figure*}%

\section{Example: Harmonic oscillator coupled to a single pseudomode} \label{sec:ho_example}
We demonstrate the thermalization behavior discussed above by considering a single harmonic oscillator as an example system,
$H_S = \hbar\omega_0 b^\dag b$, which is coupled to an environment represented by a pseudomode
(with creation and annihilation operators $a^\dag$ \& $a$) in one of three ways:
\begin{itemize}
\item $H_{SP} = \hbar g \left(b a^\dag + b^\dag a\right)$, as in~\eqref{eq:Ham_RWA}, with the residual environment at $T > 0$.
\item $H_{SP} = \hbar g \left(b + b^\dag\right) \left(a + a^\dag\right)$, as in~\eqref{eq:Ham_noRWA}, with the residual $T > 0$.
\item $H_{SP} = \hbar g \left(b + b^\dag\right) \left(a + a^\dag\right)$, as above, but with a residual $T = 0$.
\end{itemize}
This section is concerned mostly with the resonant case $\omega_0 = \omega_P$.
The off-resonant case and its effects will be discussed in more detail in Sec.~\ref{sec:nh}.

The steady states were computed exactly using
third-quantization methods~\cite{Prosen2010} (see App.~\ref{app:third_quantization} for details),
which are closely related to covariance-matrix methods~\cite{Serafini}.
These results, for $\omega_0 = \omega_P$, can be summarized as follows
(non-RWA results are shown in Fig.~\ref{fig:combined}):
\begin{itemize}
    \item The RWA model thermalizes to the target temperature for any combination of $\gamma$ and $g$.
    \item The $x$-$x$ model for $T = 0$ thermalizes to the $T_\text{target}$ state for small $g$ only when the
        pseudomode's damping rate $\gamma$ satisfies the condition~\eqref{eq:gamma_fixed}.
    \item The $x$-$x$ model for $T = T_\text{target}$ thermalizes to the $T_\text{target}$ state
        only if $g$ and $\gamma$ are both very small, as confirmed by the analytical result~\eqref{eq:betamax} given below.
\end{itemize}
More specifically, if the RWA is applied, the combined system--pseudomode steady state is found to be---%
for arbitrary $\gamma > 0$ and $g > 0$ (and for both $\omega_P = \omega_0$ and $\omega_P \neq \omega_0$):
\[Z^{-2}\, e^{- \frac{\beta \omega_P}{\omega_0} \hbar \omega_0 b^\dag b} \otimes e^{- \beta \hbar \omega_P a^\dag a},\]
where $Z = \left(1 - e^{-\beta \hbar \omega_P}\right)^{-1}$.
That is, the effective temperature experienced by the system mode is
$\beta' = \beta \omega_P / \omega_0$, or $T' = \omega_0 T / \omega_P$ (cf.~\eqref{eq:Btemp}).

However, if the RWA is not applied, the steady state becomes more complicated.
The system oscillator is, in general, found to be in a squeezed thermal state
$\mathfrak{s}(\zeta) \rho_{\tilde{\beta}} \mathfrak{s}^\dag(\zeta)$
with $\mathfrak{s}(\zeta) \coloneqq e^{\frac{1}{2} \left(\zeta^\ast b^2 - \zeta \left(b^\dag\right)^2\right)}$
and $\rho_{\tilde{\beta}} = \left(1 - e^{-\tilde{\beta}\hbar\omega_P}\right) e^{- \tilde{\beta}\hbar \omega_P b^\dag b}$.
The effective temperature $\tilde{\beta}$ of this steady state for $\omega_0 = \omega_P$
is given by (see App.~{\ref{app:squeezing}} for details)
\begin{equation}
\tilde{\beta} \hbar \omega_0
    = \ln(\frac{\sqrt{\frac{2 g^2 {g_c}^2 \nu^2}{{g_c}^2 - g^2} + \nu^4} + \frac{{\omega_0}^2}{2N+1}}{
                \sqrt{\frac{2 g^2 {g_c}^2 \nu^2}{{g_c}^2 - g^2} + \nu^4} - \frac{{\omega_0}^2}{2N+1}}),
\label{eq:beta_tilde}
\end{equation}
where $\nu^2 \coloneqq {\omega_0}^2 + \frac{\gamma^2}{8}$ and the critical value $g_c$ is:
\begin{equation}
g_c \coloneqq \frac{1}{2} \sqrt{\omega_0 \omega_P + \frac{\omega_0}{\omega_P}\frac{\gamma^2}{4}}.
\label{eq:g_crit}
\end{equation}
The squeezing factor $\zeta$ of the steady state is (for $\omega_0 = \omega_P$)
\begin{equation}
\zeta = -\frac{1}{4}\ln(\frac{2g^2 {g_c}^2}{\nu^2 ({g_c}^2 - g^2)} + 1).
\label{eq:squeezing}
\end{equation}
Note that, as $g \to 0$, $\zeta$ also decays to zero, indicating that the state becomes purely thermal in this limit,
and
\begin{equation}
\tilde{\beta} \to \frac{1}{\hbar \omega_0} \ln(1 + \frac{16{\omega_0}^2}{\gamma^2 + 2 N \left(8 {\omega_0}^2 + \gamma^2\right)}),
\label{eq:betamax}
\end{equation}
which corresponds exactly to the target temperature $\beta$ if~\eqref{eq:gamma_fixed} is fulfilled.

As the system--bath coupling $g$ approaches the value $g_c$,
the average particle number $\expval{b^\dag b}$ in the $x$-$x$-coupled non-RWA models increases up to infinity
(accordingly, $\tilde{\beta} \to 0$, but also $\zeta \to -\infty$).
Any $g$ above that threshold causes one of the normal modes of the combined system--pseudomode supersystem
to effectively obtain an imaginary frequency, leading to unphysical steady states.
In other words, if both the system and the pseudomode are represented by unbounded harmonic oscillators,
then the system--bath coupling may never exceed $g_c$---%
but if the harmonic oscillators are truncated, then this bound may be exceeded,
albeit with potentially unpredictable results.

In App.~{\ref{app:counterterms}}, we investigate the effects of including counterterms
in the spirit of the Caldeira--Leggett model~{\cite{Caldeira1983AnnPhys}} and show that,
although including such counterterms increases the value of $g_c$ and therefore delays the onset of the divergent behavior,
it does not eliminate the divergence and makes no qualitative difference in the situations considered here.

We point out that the limit of $\gamma \to \infty$ produces increasingly flat BCFs, which, in a sense,
represent an increasingly Markovian environment.
However, due to the nature of the Lorentzian BCF, this flatness extends across the entire real line.
In other words, although $\gamma \to \infty$ can be expected to produce quasi-Markovian dynamics,
bath-induced excitation and relaxation processes obtain equal weights in this limit,
which suggests equilibration to a thermal state whose effective temperature grows without bound.
Indeed, taking the limit $\gamma \to \infty$ in~\eqref{eq:beta_tilde}, \eqref{eq:g_crit} and \eqref{eq:squeezing} shows that
\[g_c \to \infty, \quad \zeta \to 0, \quad \tilde{\beta} \to 0 \iff T_\text{eff} \to \infty,\]
confirming that the quasi-Markovian limit $\gamma \to \infty$ suppresses nonequilibrium effects
and causes equilibration to a thermal state of diverging effective temperature for an increasingly large region of $g < g_c$.
This stands in contrast to the relatively small-$\gamma$ cases discussed above that exhibit
thermalization only in the limit of small $g$, a limit that can be interpreted as minimizing the effects
of the pseudomode's non-Markovianity on the system.

A quintessential example of such a quantum harmonic oscillator interacting strongly with a non-Markovian environment
is a vibrational degree of freedom of a molecule in a liquid.
Here, interactions with the vibrations of other molecules in the densely-packed condensed phase result in strong coupling.
In particular, if such a system is also strongly coupled to light (giving rise to vibrational polaritons~\cite{Xiong2023ACR}),
then strong-coupling-induced deviations from thermal behavior may play a crucial chemical role~\cite{Du2021JCP,Xiong2023ACR}.
Moreover, harmonic oscillators may be natural extensions of two- or three-level system heat engines~\cite{Albarelli2025QST,Weber2026NJP}
to continuous-variable (or hybrid) systems such as trapped ions~\cite{Sun2025NatComms,So2025NatComms}.


\section{Non-Hermitian pseudomodes and flat effective-temperature profiles} \label{sec:nh}
Although the preceding analysis has established that pseudomodes may provide detailed balance for a
specific transition frequency $\omega$, it is clear from Fig.~\ref{fig:basic_bcfs}\textit{\textbf{c}}%
---and the analysis given in the weak-coupling section above---%
that the temperature that the system will \enquote{see} depends heavily on the frequency of the transition that is being mediated.
This is an undesirable artifact of the method, since one would typically expect a physical
bath to provide the same temperature regardless of the Bohr frequency in question.
In particular, one could contemplate whether the frequency-dependent temperature is the reason
for the departure from the thermal state in the non-RWA models.

Thus, to further investigate this phenomenon and construct a pseudomode environment that affords a flatter
temperature profile, we turn to combinations of several pseudomodes.
A central issue to be dealt with in this endeavor are the infamous fat tails exhibited by
the Lorentzian BCFs resulting from uncoupled pseudomodes, which, in this context,
\enquote{leak} into regions far away from their resonant frequency, resulting in unwanted pump-like effects~\cite{Lednev2024PRL}.
As has been noted in the past~\cite{Lambert2019NatComm,Luo2023PRXQuantum,Lambert2024PRR},
this issue can also be interpreted as arising from neglecting Matsubara modes and
is difficult to avoid with only positive pseudomodes, but can be alleviated by including non-Hermitian
pseudomodes~\cite{Lambert2019NatComm,Pleasance2020PRR,Menczel2024PRR}.
Non-Hermitian pseudomode contribute to the net BCF as negative Lorentzians.
However, one should note that BCFs that exhibit net-negative values (and thus negative rates) are undesirable---%
\textit{in nuce}, such BCFs are known to break complete positivity of the system's dynamics~\cite{Pleasance2020PRR, Menczel2024PRR, Lambert2019NatComm}.
Because of this, we use a construction that tunably suppresses the fat tails of the Lorentzians
while still guaranteeing a net non-negative BCF for every $\omega$ (see App.~\ref{app:nonhermitian} for details):
If one combines a standard pseudomode,
featuring some damping rate $\gamma_j$, coupling strength $g_j$ and frequency $\omega_{P,j}$,
with a non-Hermitian counterpart with the corresponding quantities
$\gamma_j' > \gamma_j$, $g_j' = i\sqrt{\alpha_j} g_j$ (where $\alpha_j > 0$) and $\omega_{P,j}' = \omega_{P,j}$,
one can show that for any $\alpha_j$ up to and including $\gamma_j / \gamma_j'$,
the net BCF resulting from the combined resonant Hermitian/non-Hermitian pseudomodes stays non-negative everywhere.
If $\alpha_j$ is kept fixed to the maximum value $\alpha_j = \gamma_j / \gamma_j'$,
then the tail suppression may be tunably increased by decreasing the value of $\gamma_j'$.

It should be noted that net-positive BCFs satisfying detailed balance can always be associated
to harmonic baths in thermal equilibrium.
Thanks to this equivalence, system dynamics generated by combinations of Hermitian/non-Hermitian pseudomodes
with flat effective-temperature profiles are guaranteed to be physical.
Furthermore, even if the $T_\text{eff}(\omega)$ generated by such a multi-mode model
is not constant, any net-positive BCF that satisfies
$\Gamma(\abs{\omega}) \geq \Gamma(-\abs{\omega})$ for every $\omega \in \mathbb{R}$
is equivalent to a (nonequilibrium) harmonic environment in a Gaussian state as in~\eqref{eq:Gamma_J}
if the inverse temperature $\beta$ inside of $N(\abs{\omega})$ is allowed to also depend on $\omega$.

\begin{figure}[t]
\includegraphics{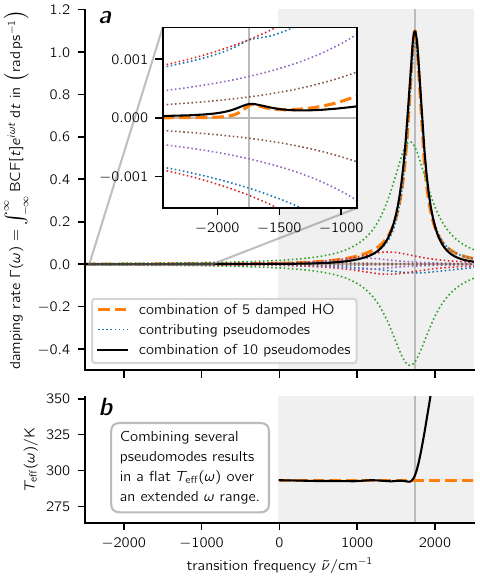}
\caption{The Fourier-transformed BCFs resulting from a specific
combination of five pseudomodes with Hermitian (H) coupling plus five pseudomodes with non-Hermitian (NH) coupling,
all under $x$-$x$ coupling~{\eqref{eq:Ham_noRWA}} to the system.
\textit{\textbf{a}}:~%
The black line represents the combined BCF resulting from the five H/NH pseudomode pairs,
which are shown as pairs of thin, colored dotted lines:
The positive-valued component of each pair is given by the H pseudomode,
whereas the positive-valued component corresponds to its NH counterpart.
The orange dashed line represents the corresponding combination of five Ohmically damped HOs
(as given by $J_\text{ref}(\omega)$~{\eqref{eq:Jref}})
at the same frequencies as the pairs of pseudomodes.
\textit{\textbf{b}}:~The effective temperature as given by~{\eqref{eq:T_eff}}.
Compared to the single-pseudomode BCFs as shown in Fig.~{\ref{fig:basic_bcfs}\textit{\textbf{c}}},
the combination of several pseudomodes clearly allows for a nearly constant effective temperature
over a much wider range of transition frequencies.
The full table of parameters is given in App.~{\ref{app:nonhermitian}}.
}
\label{fig:nh_bcfs}
\end{figure}

\begin{figure}
\includegraphics{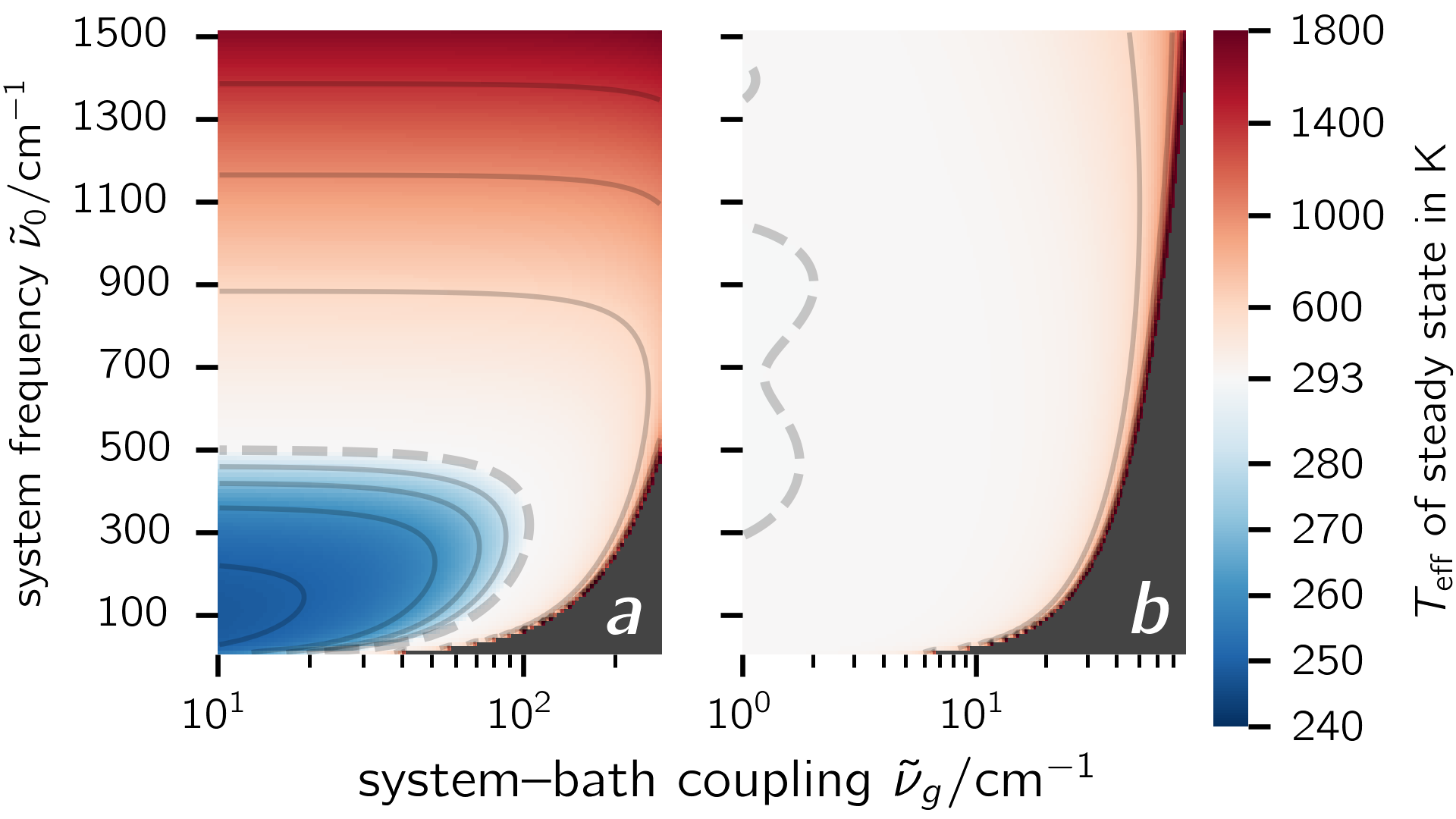}
\caption{The effective temperature $T_\text{eff}$ of the steady state of a harmonic oscillator with variable frequency $\omega_0$
when (\textbf{\textit{a}}) coupled to a single non-RWA pseudomode with $T = 0$ and $\gamma =\gamma_\text{fixed}$
as in~\eqref{eq:gamma_fixed},
or when (\textbf{\textit{b}}) coupled to the 10-pseudomode system shown in Fig.~\ref{fig:nh_bcfs}.
Note that the colorbar has differing slopes for
$T_\text{eff} > T_\text{target}$ and $T_\text{eff} < T_\text{target}$.
The dashed contour line marks $T_\text{eff} = T_\text{target} = \SI{293}{\K}$,
and the gray area is the region in which $g > g_c$ from~\eqref{eq:g_crit}.
\textbf{\textit{a}}:~%
Under the influence of a single pseudomode, $T_\text{eff}$ depends on the system frequency $\omega_0$
even for a small $g$. In the limit of small $g$, $T_\text{eff} = T_\text{target}$ is fulfilled only
when the resonance criterion $\omega_0 = \omega_P$ is satisfied.
Note that the steady state is also squeezed whenever $g$ is non-negligible, as given in~\eqref{eq:squeezing}.
\textbf{\textit{b}}:~%
When coupled to the fine-tuned combination of ten pseudomodes,
the system's $T_\text{eff}$ becomes constant for weak coupling $g$,
in accordance with the $T_\text{eff}$ of~\eqref{eq:T_eff} as computed from the BCF (also shown in Fig.~\ref{fig:nh_bcfs})---%
unlike in the single-pseudomode situation shown in \textbf{\textit{a}},
which results in a strong frequency dependence even for small $g$.
For the smallest displayed coupling of $\tilde{\nu}_g = \SI{1}{\per\cm}$, $T_\text{eff}$ ranges from \SI{292.7}{\K}
to \SI{296.5}{\K}. 
}
\label{fig:double_2d}
\end{figure}

A combination of five Hermitian plus five non-Hermitian pseudomodes that provides
a flat temperature profile using this technique in combination with
\enquote{pseudo-Schrödinger equations}~\cite{Lambert2019NatComm,Luo2023PRXQuantum} is shown in Fig.~\ref{fig:nh_bcfs}.
The parameters used in this specific construction are given in Tab.~\ref{tab:nh_params} of App.~\ref{app:nonhermitian}.
A crucial consequence of the flat temperature profile is that this multi-pseudomode model
should thermalize a system with arbitrary Bohr frequencies up to roughly \SI{1800}{\per\cm}
to the same target temperature of \SI{293}{\K} (in the limit of weak coupling, and assuming a unique steady state),
unlike the single-pseudomode BCFs shown in Fig.~\ref{fig:basic_bcfs},
which are only tuned to the target temperature at a specific transition frequency and will produce
different steady-state temperatures for different transition frequencies.
This behavior is confirmed in Fig.~\ref{fig:double_2d} for a system composed of a single harmonic oscillator---%
note that in this Figure, the nominal multi-pseudomode coupling parameter $g$ is scaled in such a way
that a given value of $g$ here produces the same net resonant damping rate $\Gamma(\omega_0)$
as a given single-pseudomode $g$, which makes the coefficient $g$ comparable across the different models.

We emphasize that the corrections due to the non-Hermitian pseudomodes impact not just the steady states
but the underlying BCF and, therefore, any dynamical property that depends on the BCF:
Rather than merely correcting steady-state populations,
the additional non-Hermitian pseudomodes thus act to compensate the KMS violations introduced by the Lorentzian representation.
Viewed from this perspective, the non-Hermitian modes may be interpreted as partially restoring the thermal symmetry
of the original environment, which provides an explanation for their improved thermalization behavior.

\section{Conclusions}\label{sec:conclusions}
We have investigated the thermalization and detailed-balance behavior of pseudomode models,
a class of models known to be capable of exactly reproducing dynamics of open quantum systems
beyond the perturbative regime, but which require the bath correlation functions
to be representable as a combination of Lorentzians.
In general, thermalization and detailed balance can only be expected to hold true in the weak system--bath coupling case,
and we investigate this case by approximating the pseudomode dynamics by Markovian master equations,
which reveals various conditions under which pseudomode dynamics will or will not exhibit the desired behavior.
Surprisingly, in many situations, even perturbatively weak coupling will \emph{not} lead to detailed balance
and thermalized (Gibbs) steady states, demonstrating the violation of the KMS structure of the original thermal environment
introduced by the effective pseudomode modeling.
Our analysis of system--bath coupling  with and without the RWA extends recent weak-coupling results on detailed balance
beyond the RWA~\cite{Scandi2026PRX} to the non-perturbative regime,
as our analytical results are supported by numerical, nonperturbative pseudomode calculations.

To construct a thermodynamically consistent effective pseudomode environment at a specific, well-defined temperature,
one of the following strategies could be employed:
First, one could apply a RWA while also ensuring that the frequency of the pseudomode is close to the range of Bohr frequencies whose transitions it mediates.
This approach is stable over a range of system--pseudomode coupling strengths, but is restricted to a single Bohr frequency
and requires justifying the use of the RWA.
If one does not want to apply a RWA:
\begin{itemize}
    \item Choose $\gamma$ according to~\eqref{eq:gamma_fixed}. This results in the correct temperature only for a single Bohr frequency.
    \item Combine fine-tuned several pseudomodes, likely including non-Hermitian pseudomodes. This can result in approximately attaining $T_\text{target}$ over a wider range.
    \item Combine several fine-tuned coupled pseudomodes in the style of Ref.~\cite{Lednev2024PRL}. This can also result in attaining $T_\text{target}$ over a wide range.
\end{itemize}
We reiterate that the non-RWA cases, even under the conditions stated above,
will thermalize only for weak system--pseudomode coupling $g$,
and exhibit increasingly athermal behavior as $g$ approaches $g_c$~\eqref{eq:g_crit}.
Such thermodynamically consistent non-RWA pseudomode models that thermalize to the
Gibbs state for weak coupling are excellent platforms for investigating the transition from
(weak-coupling) equilibrium thermalization to (strong-coupling) nonequilibrium steady states in realistic settings
that reproduce mean-force Gibbs states~\cite{Miller2018,Trushechkin2022AVS}.
Several past investigations of quantum thermodynamics using pseudomodes have
noted the nontrivial relationship between the pseudomode's \enquote{residual environment}
temperature and the temperature experienced by the system~\cite{Lambert2019NatComm,Lambert2024PRR,Lednev2024PRL},
but some studies have not taken this into account, which probably quantitatively skews the results~\cite{Albarelli2025QST},
while some other studies have likely avoided most of the issue (whether consciously or not) by
applying the system--bath RWA~\cite{Wu2022EPL,Lacerda2023PRB,Alamo2024PRE}.
Furthermore, future work would need to consider in more detail to what extent the results presented here
apply to the recently introduced fermionic counterparts of pseudomodes
known as pseudofermions~\cite{cirio_pseudofermion_2023}.

We note that reaction coordinate (RC) mapping~\cite{Nazir2018}, a numerical procedure similar to pseudomode methods,
typically employ a \enquote{global} (Bloch--Redfield/Born--Markov or GKSL) master equation to model the residual environment
and, therefore, might not suffer from the same artificial-pumping artifacts discussed in this manuscript%
~\cite{Iles-Smith2014PRA,Iles-Smith2016JCP,Anto-Sztrikacs2021NJP,Ivander2022PRE}
(naturally, deviations from the system Gibbs state are still to be expected for non-weak coupling
where the mean-force Gibbs state emerges instead~\cite{Trushechkin2022AVS})---%
although it should be noted that the use of the nonsecular Bloch--Redfield equation may, in unfavorable circumstances,
cause different non-equilibration artifacts compared to the secular global GKSL equations.
Additionally, we point out that determining the jump operators used in such global master equations
as often employed in RC mappings requires diagonalizing the Hamiltonian of the system including its reaction coordinates,
which may become unfeasible for complicated systems, whereas the pseudomode approach discussed in the present manuscript
uses much simpler \enquote{local} jump operators in the pseudomode dissipators~\eqref{eq:L_post}
that do not necessitate diagonalization of the system or any part of it.
Moreover, previous studies~\cite{Rivas2010NJP} indicate that such global master equations may result in artifacts in certain
coupling regimes;
in contrast, the controlled approach employed by pseudomodes avoids such artifacts
if the pseudomodes' BCFs match the original environments' BCFs sufficiently well~\cite{Tamascelli2018PRL}.

In conclusion, for non-weak coupling, pseudomode models do not generally thermalize to the system Gibbs state---%
which is to be anticipated, given that non-weak coupling exceeds the typical confines of equilibrium thermodynamics
and is therefore expected to result in a mean-force Gibbs state~\cite{Miller2018,Trushechkin2022AVS}.
However, we believe that a thermodynamically consistent model should reproduce known equilibrium thermodynamics
when the system--bath coupling of a given model is reduced to infinitesimally weak levels.
We have demonstrated that thermalization itself may constitute a particularly sensitive diagnostic
of the thermodynamic fidelity of an effective bath representation in the weak-coupling limit.
This may reveal deficiencies that remain essentially invisible at the level of transient real-time dynamics,
but which could nevertheless become visible through violations of detailed balance and asymptotic thermodynamic behavior.
Therefore, our analysis should provide an important foundation for future studies of quantum thermodynamics
using pseudomodes across all coupling regimes, as it provides a framework to determine
the thermodynamic self-consistency of a given pseudomode model.
Furthermore, the principles applied in our analysis of the thermodynamic consistency of pseudomodes
applies to any method that relies on an effective bath representation:
Any \emph{exact} representation of a thermal bath will cause a weakly-attached system to thermalize,
but all practical open quantum system methods introduce some level of (truncation) artifacts to this representation,
meaning that our analysis conceptually extends to a wide variety of methods ranging from
collision models~\cite{ciccarello_quantum_2022, lacroix_making_2025, christensen_ancilla-train_2025}
to hierarchical equations of motion~\cite{tanimura_numerically_2020, Lacroix2026} and chain mappings~\cite{chin_exact_2010,Lacroix2026}.

\section*{Acknowledgments}
We thank Dario Tamascelli, Johannes Feist, Mark Mitchison, Gabriela Wójtowicz, Giovanni Di Meglio, and Nicola Lorenzoni for interesting discussions and comments.
We acknowledge support provided by
the ERC Synergy grant HyperQ (Grant No.~856432),
the German Federal Ministry of Science (BMFTR) under the projects SPINNING (Grant No.~13NI6215) and PhoQuant (Grant No.~13N16110),
and the Volkswagen Foundation -- 0200187. 
\bibliographystyle{quantum}
\bibliography{refs.bib}

\onecolumngrid
\appendix
\section{Thermal pseudomode bath correlation functions}\label{app:bcf_basics}
We shall compute the bath correlation function (BCF) of such a pseudomode for non-zero temperatures
and any time $t \in \mathbb{R}$ using the terminology of Ref.~\cite{Tamascelli2018PRL}.

A technical point to be noted: To be able to follow the derivation of Ref.~\cite{Tamascelli2018PRL},
the auxiliary or residual environment $\tilde{E}$ that Tamascelli \textit{et al.} employ should
still be kept at $T = 0$ and the effect of temperature should instead be included
by adding terms to the interaction $R - \tilde{E}$ and setting the relative rates in a way
that reproduces $R$ being at a non-zero temperature.
This is effectively the same as applying a thermalized spectral density (T-SD) on the $R - \tilde{E}$ interaction~\cite{Tamascelli2019PRL}.
However, since the effect of the T-SD on $R$ and thus on $S$ is the same as a true $T > 0$ environment,
we refer to this effective temperature as the temperature of $\tilde{E}$ in the following to avoid unnecessary confusion. 

One can easily verify that the thermal state
\[\rho_\theta = \left(1 - e^{-\beta\hbar\omega_R}\right) \sum_{k = 0}^\infty e^{- k \beta\hbar\omega_R} \dyad{k}\]
is a steady state of $\mathcal{L}_R$, i.e., $\mathcal{L}_R[\rho_\theta] = 0$.
The bath correlation functions of the operators $a$ and $a^\dag$ may then be computed
essentially by generalizing equation~(18) of Ref.~\cite{Tamascelli2018PRL}:
\[C^L_{aa^\dag}(t) = \Tr_R\fleft\{a e^{\mathcal{L}_R t}\fleft[a^\dag \rho_\theta\fright]\fright\}\quad \forall\ t > 0.\]
Straightforward calculations show that
\begin{align*}
\mathcal{L}_R\fleft[a^{\dagger} \rho_\theta \fright]
    &= \left(-i\omega_R - \frac{\gamma}{2}\right) (a^{\dagger}\rho_\theta)
\shortintertext{and therefore}
e^{\mathcal{L}_R t}\fleft[a^\dag \rho_\theta \fright]
    &= e^{\left(-i\omega_R - \frac{\gamma}{2}\right) t} \left(a^\dag \rho_\theta\right) \quad \forall\ t > 0.
\end{align*}
The same procedure may be applied to $e^{\mathcal{L}_R t}\fleft[a \rho_\theta \fright]$.
Then the thermal bath correlation functions for $t > 0$ are easily computed:
\begin{align*}
C^L_{aa^\dag}(t) &= \Tr_R\fleft\{a e^{\mathcal{L}_R t}\fleft[a^\dag \rho_\theta\fright]\fright\}
    = e^{\left(-i\omega_R - \frac{\gamma}{2}\right) t} (N + 1),\\
C^L_{a^\dag a}(t) &= \Tr_R\fleft\{a^\dag e^{\mathcal{L}_R t}\fleft[a \rho_\theta\fright]\fright\}
    = e^{\left(i\omega_R - \frac{\gamma}{2}\right) t} N,
\end{align*}
and $C^L_{a^\dag a^\dag}(t) = 0 = C^L_{aa}(t)$, as well as (by linearity),
\[
C^L_{xx}(t) \coloneqq
    \Tr_R\fleft\{\left(a + a^\dag\right) e^{\mathcal{L}_R t}\fleft[\left(a + a^\dag\right) \rho_\theta\fright]\fright\}\\
    = e^{\left(i\omega_R - \frac{\gamma}{2}\right) t} N + e^{\left(-i\omega_R - \frac{\gamma}{2}\right) t} (N + 1) \quad \forall\ t > 0.
\]

Next, note that the results given in Ref.~\cite{Tamascelli2018PRL} only apply to positive times $t > 0$,
but may be generalized to $t < 0$ by using the \enquote{out} field instead of the \enquote{in} field
in the derivation of Lemma~1 in Ref.~\cite{Tamascelli2018PRL}.
Essentially, this means replacing equation~1 in the supplemental material of Ref.~\cite{Tamascelli2018PRL}
by a generalization of~(5.3.21) of Ref.~\cite{GardinerZoller}, i.e., the starting point of Lemma~1 for $t < 0$ is
\begin{multline*}
\dv{t} \hat{O}_{SR}(t) = \frac{i}{\hbar} \comm{\hat{H}_{SR}(t)}{\hat{O}_{SR}(t)}
    + \sum_{j=1}^\ell \comm{\hat{O}_{SR}(t)}{\hat{L}^\dag_{SR}(t)}
        \left(\frac{\gamma_j}{2} \hat{L}_{R,j}(t) - \sqrt{\gamma_j}\hat{b}_\text{out}(t,j)\right)\\
    - \sum_{j=1}^\ell \left(\frac{\gamma_j}{2} \hat{L}^\dag_{R,j}(t) - \sqrt{\gamma_j}\hat{b}^\dag_\text{out}(t,j)\right)
        \comm{\hat{O}_{SR}(t)}{\hat{L}_{SR}(t)}.
\end{multline*}
One then finds that the dynamics of the pseudomode for $t < 0$ are given by a modified version $\bar{\mathcal{L}}_R$
of the Lindbladian of~(1) in the main text, but in which $\gamma \mapsto -\gamma$.
It follows that for $t < 0$
\[C^L_{aa^\dag}(t)
    = \Tr_R\fleft\{a e^{\bar{\mathcal{L}}_R t}\fleft[a^\dag \rho_\theta\fright]\fright\}
    = e^{\left(-i\omega_R + \frac{\gamma}{2}\right) t} (N + 1),\]
and thus,
\begin{equation}
C^L_{aa^\dag}(t)
    = e^{-i\omega_R t - \frac{\gamma}{2} \abs{t}} (N + 1) \quad \forall\ t \in \mathbb{R}. \label{eq:BCF_aadag}
\end{equation}
Similarly,
\begin{equation}
C^L_{a^\dag a}(t)
    = e^{i\omega_R t - \frac{\gamma}{2} \abs{t}} N \quad \forall\ t \in \mathbb{R}, \label{eq:BCF_adaga}
\end{equation}
and, by extension and for all $t \in \mathbb{R}$,
\[
C^L_{xx}(t)
    = e^{-\frac{\gamma}{2} \abs{t}} \left(N e^{i\omega_R t} + (N + 1) e^{-i\omega_R t}\right)\\
    = e^{-\frac{\gamma}{2} \abs{t}} \left(\coth(\frac{\beta \hbar \omega_R}{2}) \cos(\omega_R t) - i\sin(\omega_R t)\right).
\]
The reservoir frequency $\omega_R$ corresponds to $\omega_P$ in the main text ($P$ standing for pseudomode),
or, when several pseudomodes are involved, $\omega_{P,j}$.
Furthermore, for clarity, we denote the correlation functions as $\expval{A(t) B(0)}$ instead of $C^L_{AB}(t)$ in the main text.

\section{When is the steady state of a pseudomode model unique?}\label{app:uniqueness}
Using Corollary~5.4 of Ref.~\cite{Zhang2024JPA} (the \enquote{extension of Frigerio's second theorem}),
we determine conditions for the irreducibility of the system including the pseudomode.
Since a pseudomode is an infinite-dimensional system, Theorem~3.3 of Ref.~\cite{Zhang2024JPA} tells us that
if the combined system is irreducible, then the steady state is unique if it exists.

A general uncoupled-pseudomode model is given by the generalization of~\eqref{eq:rho_SP} and~\eqref{eq:L_post} of the main text:
\begin{equation}
\pdv{t} \rho_{SP}
    = -\frac{i}{\hbar} \comm{H_S + H_{SP} + H_P}{\rho_{SP}} + \mathcal{D}[\rho_{SP}]
\label{eq:full_drho_dt}
\end{equation}
where $H_P = \sum_j \hbar\omega_{P,j} {a_j}^\dag a_j$ and
\begin{equation}
\mathcal{D}[\rho]
    = \sum_j \left[ N_j \gamma_j \left( {a_j}^\dag \rho a_j - \frac{1}{2} \acomm{a_j {a_j}^\dag}{\rho}\right)\\
    + (N_j + 1) \gamma_j \left( a_j \rho {a_j}^\dag - \frac{1}{2} \acomm{{a_j}^\dag a_j}{\rho}\right)\right]
    \qq{with $\gamma_j > 0$ for all $j$.}
\label{eq:full_dissipator}
\end{equation}
For brevity, define the total Hamiltonian $H \coloneqq H_S + H_{SP} + H_P.$

Consequently, the set of jump operators $L_\alpha$ in the language of Ref.~\cite{Zhang2024JPA} is
\[\{L_\alpha\}_\alpha = \bigcup_j \left\{\sqrt{(N_j + 1) \gamma_j} a_j, \sqrt{N_j \gamma_j} a_j^\dag\right\}.\]
It follows that---as long as $N_j > 0$ (equivalently: $T_j > 0$)---the span of $\{L_\alpha\}_\alpha$ is a self-adjoint set
and therefore we can work with $H$ instead of Zhang and Barthel's
$K = \frac{i}{\hbar} H + \frac{1}{2} \sum_\alpha L_\alpha^\dagger L_\alpha$
in the following.

The most general possible Hermitian system--pseudomode coupling operator
that is either $x$-type or RWA-type in the pseudomode part is
\begin{equation}
H_{SP} = \hbar \sum_{j} \left[ g_j {A_j}^\dag a_j + {g_j}^\ast A_j {a_j}^\dag \right]
\qq{with $g_j \neq 0$ for all $j$.}
\label{eq:full_HSP}
\end{equation}
Naturally, if any $g_j = 0$, then the $j$-th pseudomode should simply be removed from the equations.
To relate this more general form to the examples given in the main text,
note that~\eqref{eq:Ham_noRWA} can be reproduced by setting $A_j = S_j + S_j^\dag$,
while \eqref{eq:Ham_RWA}~corresponds to $A_j = S_j$ and~\eqref{eq:Ham_zx} corresponds to $A_j = Z_j$
(all three with the additional assumption that every $g_j \in \mathbb{R}$).

To establish whether the dynamics produced by~\eqref{eq:full_drho_dt} (for an arbitrary $H_S$)
are irreducible, we shall determine the commutant of the set
\[\{H\} \cup \mathcal{A}
    \qq{where} \mathcal{A} \coloneqq \bigcup_j \left\{a_j, {a_j}^\dag\right\},\]
which, under the aforementioned conditions, is a valid representative set $\mathcal{G}$
in the sense of Corollary~5.4 of Ref.~\cite{Zhang2024JPA}.
Since any element of the commutant of $\{H\} \cup \mathcal{A}$ must also commute
with each $a_j$ and ${a_j}^\dag$ for every pseudomode Hilbert space $j$,
we can conclude that the elements of the commutant must act trivially ($\mathbbm{1}_j$) on all of these spaces
and may act non-trivially only on the system Hilbert space $\mathcal{H}_S$.
Let $X$ be the $\mathcal{H}_S$-component of an element of the commutant.
Then, necessarily:
\[\comm{X \otimes \bigotimes_j \mathbbm{1}_j}{H_S + H_{SP} + H_P} = 0
    = \comm{X}{H_S} + \hbar\sum_{j} \left( g_j \comm{X}{{A_j}^\dag} a_j + {g_j}^\ast \comm{X}{A_j} {a_j}^\dag \right),\]
where we used $\comm{\bigotimes_j \mathbbm{1}_j}{H_P} = 0$.
Now, note that $a_j$ and ${a_j}^\dag$ appear in the individual terms of the sum but, by definition,
cannot appear in the first commutator or in any term of the sum with index $k \neq j$.
Therefore, each of the commutators must vanish individually.
From this, we conclude the following theorem:
\begin{theorem} \label{thm:uniqueness}
Let a pseudomode-coupled system be governed by the master equation~\eqref{eq:full_drho_dt}
with $\mathcal{D}$ as given in~\eqref{eq:full_dissipator} with $N_j > 0$ for all $j$
and $H_{SP}$ as in~\eqref{eq:full_HSP}.
Let $\mathcal{C}$ be the following commutant set:
\begin{equation*}
\mathcal{C} = \left\{X \in \mathcal{L}(\mathcal{H}_S) \mid
    \forall\, j: \comm{X}{A_j} = \comm{X}{{A_j}^\dag} = \comm{X}{H_S} = 0\right\}.
\end{equation*}
If $\mathcal{C} = \{z \mathbbm{1} \mid z \in \mathbb{C}\}$---%
that is, if only the identity and multiples of the identity commute with all $A_j$, ${A_j}^\dag$ and $H_S$---%
then the pseudomode-coupled system described by~\eqref{eq:full_drho_dt}
either has a \emph{unique} steady state or no steady state at all.
\end{theorem}
It is easy to verify that the harmonic-oscillator system of Sec.~\ref{sec:ho_example} satisfies these criteria when $T > 0$:
Both with and without the RWA, $H_S = \hbar\omega_0 b^\dag b$. If the RWA is not applied, then there is only one $A_j = b + b^\dag$.
One then needs to determine all $X$ that satisfy
\[\comm{X}{b^\dag b} = \comm{X}{b + b^\dag} = 0.\]
One easily sees that any such $X$ must be proportional to the identity.
Combined with the fact that we previously determined the steady state by explicit construction
(see App.~\ref{app:third_quantization}),
we know that said steady state must be the only possible steady state.
An analogous argument applies to show that the steady state under the RWA is also unique.

On the other hand, in $z$-$x$-coupled cases discussed in the main text,
if the operators $Z_j$ (which represent $A_j$ in this case) are truly diagonal, then $\comm{Z_j}{H_S} = 0$.
Then any linear combination of $Z_j$ and $H_S$ would also satisfy
\[\comm{Z_j + H_S}{H_S} = \comm{Z_j + H_S}{Z_j} = 0\]
and therefore such a system \emph{cannot} be guaranteed to feature a unique steady state
(indeed, one can often easily verify that a pure-dephasing model does not have a unique steady state).
However, if $Z_j$ is only approximately diagonal, such that $\mathcal{B}_{Z_j}$ also contains nonzero elements
and thus $\comm{Z_j}{H_S} \neq 0$, then a unique steady state may again exist, depending on the specific situation.

Theorem~\ref{thm:uniqueness} is based on Corollary~5.4 in Ref.~\cite{Zhang2024JPA},
which requires the Hamiltonian $H$ to be Hermitian.
If $H$ is non-Hermitian, Corollary~5.4 does not hold, but Theorem~5.2 of Ref.~\cite{Zhang2024JPA} can still be applied,
although its application is slightly more complicated;
one can show that Theorem~5.2 applies to the non-Hermitian one-oscillator/10-pseudomode example given in Sec.~\ref{sec:nh}.

Furthermore, we note that the steady states obtained using the third-quantization procedure
for systems of harmonic oscillators are always unique, including for $T = 0$~\cite{Prosen2010}.

\section{Adding a single negative-frequency pseudomode}\label{app:negative_mode}
Consider the $T = 0$, non-RWA case with a positive-frequency pseudomode at $\omega_{P,j}$ and coupling $g_j$,
as well as a negative-frequency pseudomode at $-\omega_{P,j}$ and coupling strength $\kappa_j g_j < g_j$.
The damping rate is then
\[\Gamma_j(\omega)
    = {g_j}^2 \left[ \frac{\gamma_j}{\frac{{\gamma_j}^2}{4} + \left(\omega - \omega_{P,j}\right)^2}
        + \frac{{\kappa_j}^2 \gamma_j}{\frac{{\gamma_j}^2}{4} + \left(\omega + \omega_{P,j}\right)^2} \right].\]
The corresponding reverse rate would be
\[\Gamma_j(-\omega)
    = {g_j}^2 \left[ \frac{\gamma_j}{\frac{{\gamma_j}^2}{4} + \left(\omega + \omega_{P,j}\right)^2}
        + \frac{{\kappa_j}^2 \gamma_j}{\frac{{\gamma_j}^2}{4} + \left(\omega - \omega_{P,j}\right)^2} \right].\]
In order for (strong) detailed balance to hold, one would require
\[{\kappa_j}^2 \overset{!}{=} e^{-\beta\hbar\omega} \qand {\kappa_j}^2 \overset{!}{=} e^{\beta\hbar\omega}\]
simultaneously, which is clearly only possible in the high-temperature limit $\beta\hbar\omega \to 0$,
as mentioned in the main text.
A weak detailed-balance condition for $\omega \approx \omega_{P,j}$ is satisfied when $\kappa_j$ is chosen in such a way that
\[e^{\beta\hbar\omega_{P,j}} = \frac{{g_j}^2 \left[ \frac{\gamma_j}{\frac{{\gamma_j}^2}{4} + \left(\omega_{P,j} - \omega_{P,j}\right)^2}
        + \frac{{\kappa_j}^2 \gamma_j}{\frac{{\gamma_j}^2}{4} + \left(\omega_{P,j} + \omega_{P,j}\right)^2} \right]}{
            {g_j}^2 \left[ \frac{\gamma_j}{\frac{{\gamma_j}^2}{4} + \left(\omega_{P,j} + \omega_{P,j}\right)^2}
        + \frac{{\kappa_j}^2 \gamma_j}{\frac{{\gamma_j}^2}{4} + \left(\omega_{P,j} - \omega_{P,j}\right)^2} \right]}
    = 
    \frac{1 + {\kappa_j}^2 \frac{{\gamma_j}^2}{{\gamma_j}^2 + 16\left(\omega_{P,j}\right)^2}}{
            {\kappa_j}^2 + \frac{{\gamma_j}^2}{{\gamma_j}^2 + 16\left(\omega_{P,j}\right)^2}}
\]
is fulfilled.
This implies that
\[{\kappa_j}^2
    = \frac{1 - e^{\beta \hbar\omega_{P,j}} \frac{{\gamma_j}^2}{{\gamma_j}^2 + 16\left(\omega_{P,j}\right)^2}}{
        e^{\beta\hbar\omega_{P,j}} - \frac{{\gamma_j}^2}{{\gamma_j}^2 + 16\left(\omega_{P,j}\right)^2}}
    = e^{-\beta\hbar\omega_{P,j}} \frac{\left(4\omega_{P,j}\right)^2 + \left(1 - e^{\beta\hbar\omega_{P,j}}\right){\gamma_j}^2}{
        \left(4\omega_{P,j}\right)^2 + \left(1 - e^{-\beta\hbar\omega_{P,j}}\right){\gamma_j}^2}.\]
Since standard physics requires that $\beta\hbar\omega_{P,j} > 0$,
the factor $\left(1 - e^{\beta\hbar\omega_{P,j}}\right)$ will always be negative and hence ${\kappa_j}^2$
must also be negative to satisfy the preceding condition whenever
\[\left(4\omega_{P,j}\right)^2 + \left(1 - e^{\beta\hbar\omega_{P,j}}\right){\gamma_j}^2 < 0
    \iff {\gamma_j}^2 > \frac{\left(4\omega_{P,j}\right)^2}{e^{\beta\hbar\omega_{P,j}} - 1} = \left(4\omega_{P,j}\right)^2 N_j,\]
where $N_j$ is the average Bose--Einstein occupation number corresponding to an energy $\hbar \omega_{P,j}$.
Enforcing the condition on $\kappa_j$ in this regime necessarily leads to the use of non-Hermitian pseudomodes.


\section{Analytical steady state of a QHO damped by a single pseudomode}\label{app:third_quantization}
\subsection{Computation of second moments from \enquote{third quantization}}\label{app:second_moments}
We want to find an analytical expression for the steady state of the harmonic oscillator
coupled to a single pseudomode using third-quantization methods~\cite{Prosen2010}.
See also App.~\ref{app:recipe} for information on how to construct the matrices used in this section.
To do so, we need to solve equation~(23) of Ref.~\cite{Prosen2010} for
\[Z \eqqcolon \begin{pmatrix} Z_1 & Z_0 \\ {Z_0}^\mathrm{T} & Z_2\end{pmatrix}.\]
In our case, said equation reads:
\[\begin{pmatrix} \tilde{H} & \tilde{K}\\ \tilde{K}^\dag & \tilde{H}^\dag\end{pmatrix} Z
    + Z \begin{pmatrix} \tilde{H} & \tilde{K}^\dag\\ \tilde{K} & \tilde{H}^\dag\end{pmatrix}
= \begin{pmatrix} \tilde{K}^\dag & \tilde{N}\\ \tilde{N} & \tilde{K}\end{pmatrix},
\]
where
\[\tilde{H} = \begin{pmatrix} i\omega_0 & ig\\ ig & i\omega_P + \frac{\gamma}{2}\end{pmatrix},
    \quad \tilde{K} = \alpha \begin{pmatrix} 0 & ig\\ ig & 0\end{pmatrix}
    \qand \tilde{N} = \begin{pmatrix} 0 & 0\\ 0 & N \gamma\end{pmatrix}.\]
$\omega_0$ and $\omega_P$ are the frequencies of the system oscillator and pseudomode, respectively,
$N = \left(e^{\beta \hbar \omega_P} - 1\right)^{-1}$
and $\alpha = 0$ if the RWA is applied and $\alpha = 1$ if not.

Define the auxiliary variables $\delta = \omega_P - \omega_0$ and $\sigma = \omega_P + \omega_0$.
Then by essentially vectorizing the matrix $Z$, we obtain the following set of coupled linear equations:
\begin{multline*}
M
\begin{pmatrix}
Z_0^{(0,0)} &
Z_0^{(0,1)} &
Z_0^{(1,0)} &
Z_0^{(1,1)} &
Z_1^{(0,0)} &
Z_1^{(0,1)} &
Z_1^{(1,1)} &
Z_2^{(0,0)} &
Z_2^{(0,1)} &
Z_2^{(1,1)}
\end{pmatrix}^\mathrm{T}\\
=
\begin{pmatrix}
0&
0&
-i\alpha g&
0&
N \gamma&
0&
0&
0&
0&
i\alpha g
\end{pmatrix}^\mathrm{T}
\end{multline*}
with 
\[M = 
\begin{pmatrix}
0           & 2\alpha ig & 0        & 0         &
2i\omega_0  & 2ig       & 0         &
0           & 0         & 0         \\
0           & 0         & 2\alpha ig & 0        &
0           & 2ig       & 2i\omega_P + \gamma  &
0           & 0         & 0         \\
i\alpha g   & 0         & 0         & i\alpha g &
ig          & i\sigma + \gamma/2 & ig &
0           & 0         & 0         \\
0           & -ig       & ig        & 0         &
0           & -i\alpha g & 0        &
0           & i\alpha g & 0         \\
0           & ig        & -ig       & \gamma    &
0           & -i\alpha g & 0        &
0           & i\alpha g & 0         \\
-ig         & -i\delta + \gamma/2  & 0         & ig        &
-i\alpha g  & 0         & 0         &
0           & 0         & i\alpha g \\
ig          & 0         & i\delta + \gamma/2  & -ig        &
0           & 0         & -i\alpha g &
i\alpha g   & 0         & 0         \\
0           & 0         & -2\alpha ig & 0       &
0           & 0         & 0         &
-2i\omega_0 & -2ig      & 0         \\
0           & -2\alpha ig & 0       & 0         &
0           & 0         & 0         &
0           & -2ig      & -2i\omega_P + \gamma \\
-i\alpha g  & 0         & 0         & -i\alpha g &
0           & 0         & 0         &
-ig         & -i\sigma + \gamma/2 & -ig
\end{pmatrix}
\]
where $Z_k^{(l,m)}$ is the $(l,m)$-matrix element of the matrix $Z_k$
and the symmetric property of $Z_1$ and $Z_2$ was already taken into account
(i.e., $Z_1^{(0,1)} = Z_1^{(1,0)}$ and $Z_2^{(0,1)} = Z_2^{(1,0)}$).

Focus on the case $\alpha = 1$ (no RWA) first.
Define $m_k = \omega_k/g$, $d = m_P - m_0$, $s = m_P + m_0$, $c = \gamma/(2ig)$ and
\[D \coloneqq 2m_P \left(m_0 \left({m_P}^2 - c^2\right) - 4 m_P\right).\]
Then solving the matrix equation without the RWA results in:
\begin{align*}
Z_2 &= \frac{2N + 1}{D} \begin{pmatrix} \frac{m_P}{m_0} \left({m_P}^2 - c^2\right)
            & 2m_P - \frac{1}{2}(c + s)\left({m_P}^2 - c^2\right) \\
        2m_P - \frac{1}{2}(c + s)\left({m_P}^2 - c^2\right)
            & \left(m_P + c\right)^2\end{pmatrix},\\
Z_1 &= \frac{2N + 1}{D} \begin{pmatrix} \frac{m_P}{m_0} \left({m_P}^2 - c^2\right)
            & 2m_P - \frac{1}{2}(s - c)\left({m_P}^2 - c^2\right) \\
        2m_P - \frac{1}{2}(s - c)\left({m_P}^2 - c^2\right)
            & \left(m_P - c\right)^2\end{pmatrix},\\
Z_0 &= \frac{2N + 1}{D} \begin{pmatrix} 2 m_P \left(m_P + d\right) + \frac{1}{2}\left({m_P}^2 - c^2\right) \left(d^2 - c^2 - 2\frac{m_P}{m_0}\right)
            & -\frac{1}{2} (c + d) \left({m_P}^2 - c^2\right) - 2 m_P \\
        \frac{1}{2}\left(c - d\right) \left({m_P}^2 - c^2\right) - 2 m_P
            & {m_P}^2 - c^2\end{pmatrix}
        + \begin{pmatrix} N & 0\\ 0 & N\end{pmatrix}.
\end{align*}
We are primarily interested in the system-oscillator quantities
$Z_0^{(0,0)} = \expval{b^\dag b}$ and $Z_1^{(0,0)} = \expval{b^2}$.
Inserting the definitions of $c$ and $m_k$:
\begin{equation}
D  = \frac{2{\omega_P}^2 \left(\frac{\omega_0}{\omega_P} \left({\omega_P}^2 + \frac{\gamma^2}{4}\right) - 4g^2\right)}{g^4}
\label{eq:D}
\end{equation}
and then
\begin{equation}
Z_0^{(0,0)}
    = \expval{b^\dag b}
    = N + \left[\frac{g^2 \left({\omega_P}^2 + \frac{\gamma^2}{4}\right)}{
            2 \omega_0 \omega_P \left( \frac{\omega_0}{\omega_P} \left({\omega_P}^2 + \frac{\gamma^2}{4}\right) - 4g^2\right)}
        + \frac{1}{4 \omega_0 \omega_P} \left(\left(\omega_0 - \omega_P\right)^2 + \frac{\gamma^2}{4}\right)\right] (2N+1)
    \label{eq:z000}
.\end{equation}
This is evidently negative for $g > \frac{1}{2}\sqrt{\omega_0\omega_P + \frac{\omega_0}{\omega_P}\frac{\gamma^2}{4}}$,
as described in the main text. For the coherence-type term:
\begin{equation}
Z_1^{(0,0)}
    = \expval{b^2}
    = \frac{(2N+1) g^2 \left({\omega_P}^2 + \frac{\gamma^2}{4}\right)}{2 \omega_0 \omega_P \left( \frac{\omega_0}{\omega_P} \left({\omega_P}^2 + \frac{\gamma^2}{4}\right) - 4g^2\right)}
    =  Z_0^{(0,0)} - N - \frac{2N+1}{4 \omega_0 \omega_P} \left(\left(\omega_0 - \omega_P\right)^2 + \frac{\gamma^2}{4}\right).
    \label{eq:z100}
\end{equation}
On the other hand, if the RWA is applied ($\alpha = 0$), the solutions are much simpler:
\begin{equation}
Z_2 = Z_1 = \begin{pmatrix} 0 & 0 \\ 0 & 0\end{pmatrix},
\qquad
Z_0 = \begin{pmatrix} N & 0\\ 0 & N\end{pmatrix}.
\label{eq:zmat_rwa}
\end{equation}
However, note that this has slightly non-trivial implications:
Since $N = \left(e^{\beta \hbar \omega_P} - 1\right)^{-1}$, the second mode is indeed in a thermal state
at temperature $\beta$, but the first mode is in a thermal state with a different effective temperature.
Specifically, the steady state is given by
\[\left(1 - e^{-\beta \hbar \omega_P}\right)^2
    e^{- \frac{\beta \omega_P}{\omega_0} \hbar \omega_0 b^\dag b} \otimes e^{- \beta \hbar \omega_P a^\dag a},\]
that is, the effective temperature seen by the first mode is $\beta' = \frac{\beta \omega_P}{\omega_0}$,
or $T' = \frac{\omega_0 T}{\omega_P}$, as stated in the main text.

\subsection{First moments}
The preceding calculation computed the values of the second moments of the steady state.
We shall now show that the first moments all vanish.
Recall that $a$ is the annihilation operator of the pseudomode and $b$ is the corresponding system-oscillator operator.

Assume that the matrix $X$ as given in Eq.~(18) of Ref.~\cite{Prosen2010} is diagonalizable as $X = P \Delta P^{-1}$.
Define an auxiliary matrix $A = P^\mathrm{T} Z$.
Then, from eqs.~(24) and (28) of Prosen, we have, for any $r \in \{1,2,3,4\}$,
\begin{multline*}
P_{1,r} b \rho_\infty + P_{2,r} a \rho_\infty + P_{3,r} \rho_\infty b^\dag + P_{4,r} \rho_\infty a^\dag\\
    - \left[
            A_{r,1} \left(b^\dag \rho_\infty - \rho_\infty b^\dag\right)
            + A_{r,2} \left(a^\dag \rho_\infty - \rho_\infty a^\dag\right)
            - A_{r,3} \left(b\rho_\infty - \rho_\infty b\right)
            - A_{r,4} \left(a \rho_\infty - \rho_\infty a\right)\right] = 0,
\end{multline*}
where $\rho_\infty$ is the NESS.
Taking the trace makes the second line drop out:
\[P_{1,r} \expval{b} + P_{2,r} \expval{a} + P_{3,r} \expval{b^\dag} + P_{4,r} \expval{a^\dag} = 0.\]
Putting all of the different $r$ together, we obtain the following matrix equation:
\[\begin{pmatrix} \expval{b} & \expval{a} & \expval{b^\dag} & \expval{a^\dag}\end{pmatrix}
    P = \begin{pmatrix} 0 & 0 & 0 & 0\end{pmatrix}.\]
Since $P$ is a linear and invertible operator (since $X = P\Delta P^{-1}$), 
the zero vector is the unique solution to this equation
and the first moments of the steady states vanish.%

\subsection{Effective temperature \& squeezing} \label{app:squeezing}
From this, we can deduce that the system steady state cannot be displaced;
hence, by virtue of being Gaussian, it is necessarily at most a squeezed thermal state:
\[\rho_{\infty} = \frac{1}{Z} S(\zeta) e^{-\beta \hbar \omega_0 b^\dag b} S(-\zeta)
    \qq{with} S(\zeta) \coloneqq \exp(\frac{-\zeta \left(b^\dag\right)^2 + \zeta^\ast b^2}{2}).\]
We can compute the squeezing and temperature using the following relations:
\[\expval{b^\dag b} = \left(\sinh{\abs{\zeta}}\right)^2 + \frac{\cosh(2\abs{\zeta})}{e^{\beta \hbar \omega} - 1}
    \qand \expval{b^2} = -\frac{\zeta}{\abs{\zeta}} \sinh(2\abs{\zeta}) \left(\frac{1}{2} + \frac{1}{e^{\beta \hbar \omega} - 1}\right).\]
Conversely,
\[\zeta 
    = - \frac{1}{2} \frac{\expval{b^2}}{\abs{\expval{b^2}}} \artanh\fleft(\frac{\abs{\expval{b^2}}}{\expval{b^\dag b} + \frac{1}{2}}\fright)
    = - \frac{1}{4} \frac{\expval{b^2}}{\abs{\expval{b^2}}} \ln(\frac{\expval{b^\dag b} + \frac{1}{2} + \abs{\expval{b^2}}}{\expval{b^\dag b} + \frac{1}{2} - \abs{\expval{b^2}}})
\]
and
\[\tilde{\beta}
    = \frac{1}{\hbar\omega} \ln(\frac{\expval{b^\dag b} + \frac{1}{2}\left(1 + \cosh(2\abs{\zeta})\right)}{\expval{b^\dag b} + \frac{1}{2}\left(1 - \cosh(2\abs{\zeta})\right)})
    = \frac{1}{\hbar\omega} \ln(1 + \frac{1}{
                \sqrt{\left(\expval{b^\dag b} + \frac{1}{2}\right)^2 - \abs{\expval{b^2}}^2} - \frac{1}{2}})
.\]
For the RWA case, we know from~\eqref{eq:zmat_rwa} that
$\expval{b^\dag b} = \left(e^{\beta \hbar \omega_P} - 1\right)^{-1}$ and $\expval{b^2} = 0$,
hence we immediately see that $\tilde{\beta} = \beta \frac{\omega_P}{\omega}$ and $\zeta = 0$.

For the non-RWA case, one can now insert~\eqref{eq:z000} and~\eqref{eq:z100}
to obtain the effective temperature and squeezing for arbitrary parameters.
We present the explicit results only under the resonance condition $\omega_0 = \omega_P$:
\[\tilde{\beta}
    = \frac{1}{\hbar\omega_0} \ln(\frac{\sqrt{\nu^2 \left(\nu^2 + \frac{2 g^2 2{g_c}^2}{{g_c}^2 - g^2}\right)} + \frac{{\omega_0}^2}{2N + 1}}{
                \sqrt{\nu^2 \left(\nu^2 + \frac{2 g^2 2{g_c}^2}{{g_c}^2 - g^2}\right)} - \frac{{\omega_0}^2}{2N + 1}})
,\]
where, as in the main text, $g_c = \frac{1}{2}\sqrt{{\omega_0}^2 + \frac{\gamma^2}{4}}$ and $\nu^2 = {\omega_0}^2 + \frac{\gamma^2}{8}$, and
\[\zeta = - \frac{1}{4} \ln(1 + \frac{2 g^2 {g_c}^2}{\left({g_c}^2 - g^2\right) \nu^2}).\]
Evidently, the effective temperature experienced by the system mode depends on the pseudomode environment's temperature,
but the squeezing factor is independent of the pseudomode's temperature.
In the limit of small $g$, the effective temperature will tend to
\[\lim_{g\to 0} \tilde{\beta}
    = \frac{1}{\hbar\omega_0} \ln(\frac{\frac{\gamma^2}{8} + \frac{2N+2}{2N + 1} {\omega_0}^2 }{\frac{\gamma^2}{8} + \frac{2N}{2N + 1} {\omega_0}^2 }).\]
If we take the additional limit of $\gamma \to 0$, then this results in $\tilde{\beta} \hbar \omega_0  \to \ln((N+1)/N)$,
corresponding to the system mode reaching the same temperature as the pseudomode.
On the other hand, if $N = 0$ (i.e., $T = 0$ for the pseudomode), then
$\tilde{\beta} \hbar \omega_0  \to \ln(1 + 16 {\omega_0}^2/\gamma^2)$, which leads to the desired result if
the \enquote{matched} value of $\gamma$ is used.

\subsection{On the nature of the divergent behavior and counterterms} \label{app:counterterms}
In the absence of any damping, a system of two coupled modes of equal mass $m$ without the RWA
can be described using the following matrix:
\begin{equation}
H = \frac{m}{2} \begin{pmatrix} x_0 & x_P\end{pmatrix}
    \begin{pmatrix}
        {\omega_0}^2 & {\omega_g}^2\\
        {\omega_g}^2 & {\omega_P}^2
    \end{pmatrix}
\begin{pmatrix} x_0 \\ x_P\end{pmatrix}
+ \frac{{p_0}^2}{2m} + \frac{{p_P}^2}{2m}
    = 
\hbar \omega_0 \left(b^\dag b + \frac{1}{2}\right) + \hbar \omega_P \left(a^\dag a + \frac{1}{2}\right)
    + \hbar g \left(b^\dag + b\right) \left(a^\dag + a\right),
\label{eq:H_two_mode}
\end{equation}
where, for convenience, $g$ has been recast as a renormalized positive frequency
${\omega_g}^2 \coloneqq 2 g \sqrt{\omega_0\omega_P}$.
Diagonalizing:
\[
\begin{pmatrix}
    {\omega_0}^2 & {\omega_g}^2\\
    {\omega_g}^2 & {\omega_P}^2
\end{pmatrix}
=
\frac{g}{\sqrt{2 S(S + \delta)}}
\begin{pmatrix}
-\sqrt{\frac{S - \delta}{2 S}} & \sqrt{\frac{S + \delta}{2 S}}\\
\frac{g}{\sqrt{2 S(S - \delta)}} & \frac{g}{\sqrt{2 S(S + \delta)}}
\end{pmatrix}
    \begin{pmatrix}
        \frac{{\omega_0}^2 + {\omega_P}^2}{2} - S & 0\\
        0 & \frac{{\omega_0}^2 + {\omega_P}^2}{2} + S
    \end{pmatrix}
\begin{pmatrix}
-\sqrt{\frac{S - \delta}{2 S}} & \frac{g}{\sqrt{2 S(S - \delta)}}\\
\sqrt{\frac{S + \delta}{2 S}} & \frac{g}{\sqrt{2 S(S + \delta)}}
\end{pmatrix}
\]
where $S \coloneqq \sqrt{{\omega_g}^4 + \left(\frac{{\omega_0}^2 - {\omega_P}^2}{2}\right)^2}$
and $\delta = \frac{{\omega_0}^2 - {\omega_P}^2}{2}$.
In other words, the normal-mode frequencies are
\[\sqrt{\frac{{\omega_0}^2 + {\omega_P}^2}{2} \pm \sqrt{4 g^2 \omega_0 \omega_P + \left(\frac{{\omega_0}^2 - {\omega_P}^2}{2}\right)^2}}
.\]
Note that at $g = \frac{1}{2} \sqrt{\omega_0 \omega_P}$, the smaller of the two eigenfrequencies becomes zero,
and for $g$ larger than that critical value, one of the eigenfrequencies is imaginary.

\begin{figure*}[t]
\includegraphics{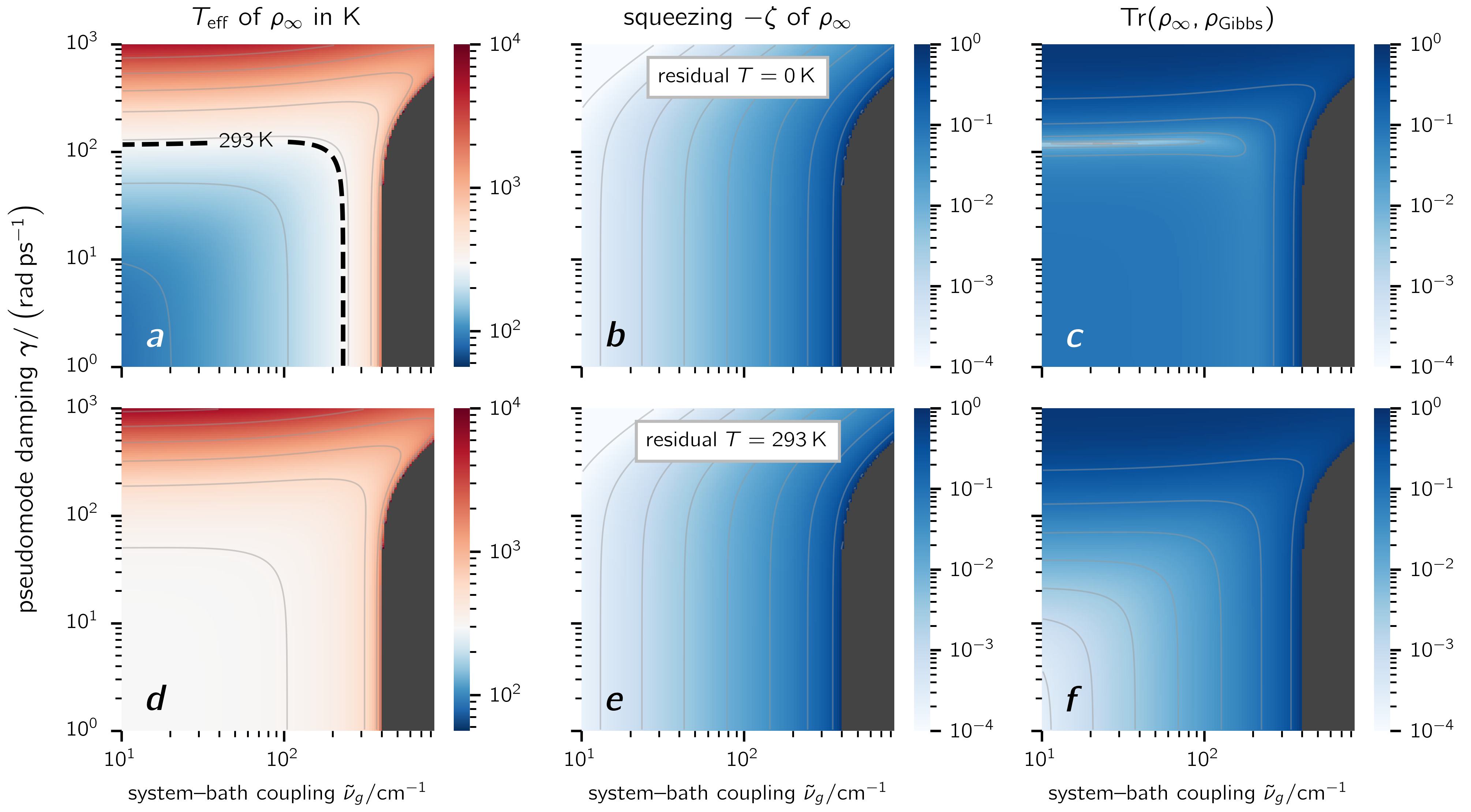}
\caption{Steady-state properties of a harmonic oscillator (at $\tilde{\nu}_0 = \SI{500}{\per\cm}$)
that is $x$-$x$-coupled~{\eqref{eq:Ham_noRWA}} to a single resonant pseudomode
\emph{including} the Caldeira--Leggett-type counterterms discussed in App.~\ref{app:counterterms}.
Apart from the counterterms, the parameters and layout of the Figure are identical to that of Fig.~\ref{fig:combined};
see the latter for details.
}
\label{fig:combined_counter}
\end{figure*}

Furthermore, one could wonder what the influence of neglecting
Lamb-shift type counterterms in the Hamiltonian~\eqref{eq:H_two_mode} is.
Compare~\eqref{eq:H_two_mode} to:
\begin{multline*}
\frac{1}{2m} \left({p_0}^2 +  {p_P}^2\right)
    + \frac{m{\omega_0}^2}{2} {x_0}^2 + \frac{m{\omega_P}^2}{2} {x_P}^2
    + \frac{m {\omega_g}^2}{2} \left(x_0 - x_P\right)^2\\
    = \frac{1}{2m} \left({p_0}^2 +  {p_P}^2\right)
        + \frac{m\left({\omega_0}^2 + {\omega_g}^2\right)}{2} {x_0}^2
        + \frac{m\left({\omega_P}^2 + {\omega_g}^2\right)}{2} {x_P}^2
        - m {\omega_g}^2 x_0 x_P.
\end{multline*}
Note that the sign of $g$ is also effectively flipped in this version, which does not change anything about the steady-state results computed previously as the latter depend only on $g^2$.
The crucial point is that the effective oscillator frequencies are modified by the inclusion of these terms;
instead of $\omega_j$, each oscillator \enquote{individually} experiences a frequency
$\sqrt{{\omega_j}^2 + {\omega_g}^2}$.
Then the frequencies of the normal modes are
\[\sqrt{\frac{{\omega_0}^2 + {\omega_P}^2 + 2{\omega_g}^2}{2}
    \pm \sqrt{{\omega_g}^4 + \left(\frac{{\omega_0}^2 - {\omega_P}^2}{2}\right)^2}}
.\]
If the modes are degenerate, i.e., $\omega_0 = \omega_P = \omega$, then the eigenmode frequencies simplify to
\[\sqrt{\omega^2 + {\omega_g}^2 \pm {\omega_g}^2} = \left\{\omega, \sqrt{\omega^2 + 2 {\omega_g}^2} \right\}.\]
In particular, this means that adding the counterterms causes the normal modes of the coupled two-mode system to always remain positive.
Nevertheless, applying the counterterm frequency shift to~\eqref{eq:D} leads to:
\[D = \frac{2\left({\omega_P}^2 + {\omega_g}^2\right) \left(\sqrt{\frac{{\omega_0}^2 + {\omega_g}^2}{{\omega_P}^2 + {\omega_g}^2}} \left({\omega_P}^2 + {\omega_g}^2 + \frac{\gamma^2}{4}\right) - 4g^2\right)}{g^4}.\]
If $\omega_0 = \omega_P$, this reduces to
\[D = 2\omega \frac{\omega\left(\omega^2 + \frac{\gamma^2}{4}\right) + 2g \left(2\omega^2 + \frac{\gamma^2}{4} - 4g^2\right)}{g^4},\]
which still becomes negative (and therefore causes divergent behavior)
for $g > g_c = \frac{1}{4}\left(\omega + \sqrt{\gamma^2 + 5\omega^2}\right)$,
albeit at a later point than without the counterterms.
Furthermore, increasing the damping $\gamma$ delays the onset of the divergent behavior somewhat.
Curiously, in the limit of infinitesimally small (but non-zero) damping $\gamma$,
the critical coupling frequency is related to the oscillator frequencies by half of the golden ratio $\phi$,
i.e., $g_c \to \phi \frac{\omega}{2}$.
The effect of including the counterterms in a setup consisting of
a single oscillator coupled to a pseudomode is demonstrated in Fig.~\ref{fig:combined_counter}.

\section{Non-Hermitian pseudomodes} \label{app:nonhermitian}
We first derive the extension of the third-quantization technique~\cite{Prosen2010}
to pseudomodes with non-Hermitian coupling in~\ref{app:nh_third_quantization}
and then demonstrate how the combination of Hermitian and non-Hermitian pseudomodes
can effectively suppress the fat tails of the Lorentzians in~\ref{app:tail_suppression}.

\subsection{Non-Hermitian third quantization} \label{app:nh_third_quantization}
For the duration of this subsection, we adopt the notation used in Ref.~\cite{Prosen2010}.

We want to expand the analysis of Ref.~\cite{Prosen2010} to allow for non-Hermitian coupling to the pseudomodes.
To that end, we need to rederive their Eq.~(14) from Eq.~(13) without assuming that $\mathbf{H} = \mathbf{H}^\dag$.
Since this change does not affect the dissipative part, we only need to consider how the first line of Eq.~(14) is changed.
Instead of their Eq.~(11), we need:
\[H = \ula^\dag \cdot \mathbf{H} \ula + \ula \cdot \mathbf{K} \ula + \ula^\dag \cdot \overline{\mathbf{J}} \ula^\dag.\]
This reduces to the standard Hermitian case if and only if $\mathbf{H} = \mathbf{H}^\dag$ \emph{and} $\mathbf{J} = \mathbf{K}$.
Inserting into their Eq.~(13):
\begin{align}
-i \hat{H}^\text{L} + i \hat{H}^\text{R}
    &= -i \left(\ula^\dag \cdot \mathbf{H} \ula + \ula \cdot \mathbf{K} \ula + \ula^\dag \cdot \overline{\mathbf{J}} \ula^\dag\right)^\text{L}
        + i \left(\ula^\dag \cdot \mathbf{H} \ula + \ula \cdot \mathbf{K} \ula + \ula^\dag \cdot \overline{\mathbf{J}} \ula^\dag\right)^\text{R} \nonumber\\
    &= -i \left[\ula_0' \cdot \mathbf{H} \ula_0 - \ula_1' \cdot \mathbf{H}^\text{T} \ula_1
            - \ula_1' \cdot \mathbf{K} \left(\ula_1' + 2 \ula_0\right)
            + \ula_0' \cdot \overline{\mathbf{J}} \left(\ula_0' + 2\ula_1\right)\right],
\label{eq:nh_Neumann}
\end{align}
where we used Eq.~(5) of Ref.~\cite{Prosen2010} along with
\[\left(\ula^\dag \cdot \mathbf{H} \ula\right)^\text{R}
    = \sum_{j, k} \mathbf{H}_{jk} \left({\hat{a}_j}^\dag \hat{a}_k\right)^\text{R}
    = \sum_{j, k} \mathbf{H}_{jk} {\hat{a}_k}^\text{R} \left({\hat{a}_j}^\dag\right)^\text{R}
    = {\ula}^\text{R} \cdot \mathbf{H}^\text{T} \left({\ula}^\dag\right)^\text{R}\]
as well as
\[\ula_1 \cdot \mathbf{H} \ula_0 - \ula_0 \cdot \mathbf{H}^\mathrm{T} \ula_1
    = \sum_{j,k} \left(a_{1, j} \mathbf{H}_{jk} a_{0,k} - a_{0, j} \mathbf{H}_{kj} a_{1, k}\right)
    = \sum_{j,k} \left(a_{1, j} \mathbf{H}_{jk} a_{0,k} - a_{1, k} \mathbf{H}_{kj} a_{0, j}\right)
    = 0
\]
and (using Eq.~(6) of Ref.~\cite{Prosen2010} and for an arbitrary $n \times n$ matrix $\mathbf{A}$)
\begin{equation}
\ula_\mu \cdot \mathbf{A} \ula_\nu'
    = \sum_{j,k} \left(a_{\mu, j} \mathbf{A}_{jk} a_{\nu,k}'\right)
    = \sum_{j,k} \left(\left(\mathbf{A}^\text{T}\right)_{kj} \left(a_{\nu,k}'  a_{\mu, j} + \delta_{\mu\nu} \delta_{jk}\right)\right)
    = \ula_\nu' \cdot \mathbf{A}^\text{T} \ula_\mu + \delta_{\mu\nu} \Tr{\mathbf{A}}.
\label{eq:ula_comm}
\end{equation}

As expected, \eqref{eq:nh_Neumann} reduces to the equation found in Ref.~\cite{Prosen2010} if
$\mathbf{H} = \mathbf{H}^\dag$ and $\mathbf{J} = \mathbf{K}$.
Given that the difference between their result and ours can be described as
$\overline{\mathbf{H}} \mapsto \mathbf{H}^\text{T}$
and $\overline{\mathbf{K}} \mapsto \overline{\mathbf{J}}$,
we can anticipate that the matrices $\mathbf{X}$ and $\mathbf{Y}$ will be modified in the following way:
\begin{align}
\mathbf{X}
    &= \frac{1}{2} \begin{pmatrix} i \mathbf{H}^\text{T} - \overline{\mathbf{N}} + \mathbf{M}
                                    & -2i \mathbf{K} - \mathbf{L} + \mathbf{L}^\text{T}\\
                                2i \overline{\mathbf{J}} - \overline{\mathbf{L}} + \overline{\mathbf{L}}^\text{T}
                                    & -i \mathbf{H} - \mathbf{N} + \overline{\mathbf{M}}
                \end{pmatrix} \label{eq:Xmat}
\shortintertext{and}
\mathbf{Y}
    &= \frac{1}{2} \begin{pmatrix} -2i \overline{\mathbf{J}} - \overline{\mathbf{L}} - \overline{\mathbf{L}}^\text{T}
                                    & 2 \mathbf{N}\\
                                2 \mathbf{N}^\text{T}
                                    & 2i \mathbf{K} - \mathbf{L} - \mathbf{L}^\text{T}
                \end{pmatrix}. \label{eq:Ymat}
\end{align}
One can insert these results into $\mathbf{S}$ of Eq.~(16) and (17) of Ref.~\cite{Prosen2010} and
ignore the components depending on $\mathbf{L}$, $\mathbf{M}$ and $\mathbf{N}$ to verify our assumption:
\begin{multline*}
i \begin{pmatrix} \ula_0 & \ula_1 & \ula_0' & \ula_1'\end{pmatrix}
    \begin{pmatrix} 0 & 0 & -\frac{i}{2} \mathbf{H}^\text{T} & i \mathbf{K}\\
                    0 & 0 & -i \overline{\mathbf{J}} & \frac{i}{2} \mathbf{H}\\
        -\frac{i}{2} \mathbf{H} & -i \overline{\mathbf{J}} & -i \overline{\mathbf{J}} & 0\\
        i \mathbf{K}    & \frac{i}{2} \mathbf{H}^\text{T} & 0 & i \mathbf{K}
    \end{pmatrix}
    \begin{pmatrix}\ula_0\\ \ula_1\\ \ula_0'\\ \ula_1'\end{pmatrix}\\
    = \ula_0 \cdot \left(\frac{1}{2} \mathbf{H}^\text{T} \ula_0' - \mathbf{K} \ula_1'\right)
        + \ula_1 \cdot \left(\overline{\mathbf{J}} \ula_0' - \frac{1}{2} \mathbf{H} \ula_1'\right) 
        + \ula_0' \cdot \left(\frac{1}{2} \mathbf{H} \ula_0 + \overline{\mathbf{J}} \left(\ula_1 + \ula_0'\right) \right)
        - \ula_1' \cdot \left(\mathbf{K} \left( \ula_0 + \ula_1'\right) + \frac{1}{2} \mathbf{H}^\text{T} \ula_1\right)\\
    = \ula_0' \cdot \mathbf{H} \ula_0 
        - \ula_1' \cdot \mathbf{H}^\text{T} \ula_1
        + \ula_0' \cdot \overline{\mathbf{J}} \left(\ula_0' + 2 \ula_1\right)
        - \ula_1' \cdot \mathbf{K} \left(\ula_1' + 2 \ula_0\right),
\end{multline*}
where we also used~\eqref{eq:ula_comm}.
This matches the right-hand side of~\eqref{eq:nh_Neumann}, as desired.

\subsection{Tail suppression by combining resonant Hermitian and non-Hermitian modes} \label{app:tail_suppression}
\begin{figure}
    \centering
    \includegraphics[width=\linewidth]{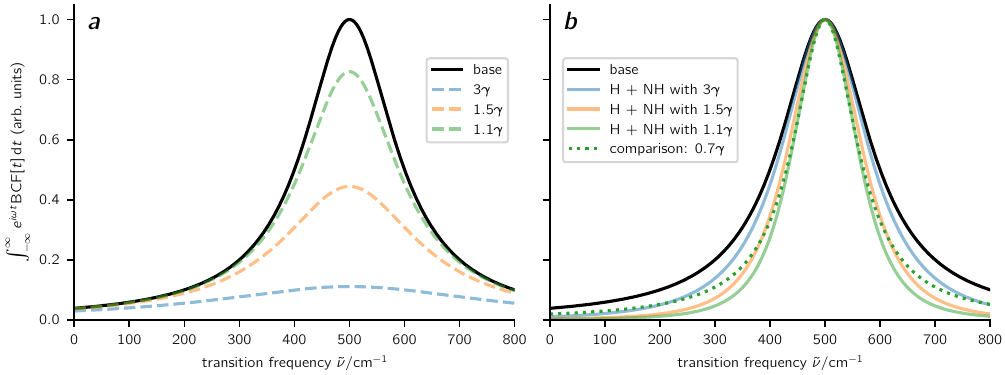}
    \caption{An illustration of the tail suppression via non-Hermitian pseudomodes
    using the maximal $\alpha = \gamma / \gamma'$ for three different values of $\gamma' \in \{3\gamma, 1.5\gamma, 1.1\gamma\}$.
    \textbf{\textit{a}}: The BCF of the base Hermitian pseudomode with $\gamma = \frac{2}{5} \omega_0$
    along with those of the three different resonant non-Hermitian pseudomodes that will serve to suppress the tails.
    The amplitudes of the non-Hermitian pseudomodes have been scaled according to $\alpha = \gamma / \gamma'$.
    \textbf{\textit{b}}: The combination of the Hermitian plus a resonant non-Hermitian pseudomode
    for the three different dampings shown in \textbf{\textit{a}}.
    The resulting combinations shown here have been rescaled according to~\eqref{eq:C_criterion}.
    The base pseudomode BCF is shown for reference, along with another simple pseudomode BCF with a \SI{30}{\percent}
    smaller width that demonstrates that the combined H+NH BCFs are qualitatively different from narrow Lorentzians.}
    \label{fig:reduced_example}
\end{figure}

Imaginary-valued system-pseudomode coupling results in a negative-valued contribution to the BCF---%
however, as described in the main text, one would like the net BCF to always remain non-negative.
It turns out that the fat tails of a single Lorentzian can be suppressed by combining
a pseudomode with a resonant non-Hermitian pseudomode while maintaining a net-positive BCF:
Let $\gamma_j$, $g_j$ and $\omega_{P,j}$ be the damping rate, coupling strength and frequency
of the standard pseudomode and $\gamma_j'$, $g_j' = i\sqrt{\alpha_j} g_j$ (where $\alpha_j > 0$)
and $\omega_{P,j}' = \omega_{P,j}$ the corresponding quantities of its non-Hermitian counterpart.
(Note that we employ a \enquote{pseudo-Schrödinger equation}~\cite{Lambert2019NatComm,Luo2023PRXQuantum}
and restrict ourselves to real damping rates, frequencies and temperatures,
allowing only the coupling $g_j$ to be either real or imaginary.)
Then the combined BCF of the Hermitian and non-Hermitian modes at zero temperature will be
(where $C > 0$ is some constant; we determine constraints on $C$ below):
\[\text{BCF}[\omega]
    = C {g_j}^2 \left(\frac{\gamma_j}{\frac{{\gamma_j}^2}{4} + \left(\omega - \omega_{P,j}\right)^2}
    - \frac{\alpha_j \gamma_j'}{\frac{{\gamma_j'}^2}{4} + \left(\omega - \omega_{P,j}\right)^2}\right).\]
In effect, adding the non-Hermitian pseudomode subtracts its BCF from the total BCF.
Accordingly, the non-Hermitian pseudomode only fulfills its role of removing the fat tails if it is
itself fat-tailed, i.e., the non-Hermitian damping must be larger than the Hermitian one:
$\gamma_j' > \gamma_j > 0$.
If we require the combined BCF to always remain non-negative, $\text{BCF}[\omega] \overset{!}{\geq} 0$,
then one easily finds that
\[\alpha_j \overset{!}{\leq}
    \frac{\gamma_j}{\gamma_j'}\, \eta
    \qq{where} \eta \coloneqq \frac{\frac{{\gamma_j'}^2}{4} + \Delta^2}{\frac{{\gamma_j}^2}{4} + \Delta^2},\]
where $\Delta \coloneqq \omega - \omega_{P,j}$.
Since $\gamma_j' > \gamma_j$ by construction, 
$\eta > 1$ for any finite value of $\Delta$, and $\eta \to 1$ as $\Delta^2 \to \infty$.
Since we require positivity for any $\omega$ and, by extension, for any $\Delta$, we thus obtain the condition
$\alpha_j \leq \gamma_j / \gamma_j'$.
Since the tail suppression is stronger the larger $\alpha_j$ is,
we will use the maximum value $\alpha_j = \gamma_j / \gamma_j'$ from here on.
It is easily verified that the combined BCF is still positive everywhere,
but we have also reduced its maximum value at $\omega = \omega_{P,j}$ by subtracting the extra mode.
We can correct for this by choosing the constant $C$ appropriately:
\begin{equation}
{g_j}^2 \frac{\gamma_j}{\frac{{\gamma_j}^2}{4} + \left(\omega_{P,j} - \omega_{P,j}\right)^2}
    = \frac{4 {g_j}^2}{\gamma_j}
    \overset{!}{=} \text{BCF}[\omega_{P,j}]
    \iff C = \frac{\gamma_j'}{\gamma_j' - \alpha_j \gamma_j}.
\label{eq:C_criterion}
\end{equation}
For the maximal $\alpha_j = \gamma_j/\gamma_j'$, this results in $C = 1/\left(1 - {\alpha_j}^2\right)$.
The tail suppression that is achieved by such a combination of Hermitian and non-Hermitian pseudomodes is illustrated in
Fig.~\ref{fig:reduced_example}.

Reintroducing a non-zero residual environment temperature of $\beta$,
we can compute the effective temperature $\tilde{\beta}$ of such a combined partially non-Hermitian
pseudomode environment at $\omega = \omega_{P,j}$ using the maximal $\alpha_j$:
\begin{equation}
e^{\tilde{\beta} \hbar\omega_{P,j}}
    = \frac{\text{BCF}[\omega_{P.j}]}{\text{BCF}[-\omega_{P.j}]}
    =  e^{\beta\hbar \omega_{P,j}} \frac{%
    \left(1 + r_j\right) \left(1 + {\alpha_j}^2 r_j\right) + e^{-\beta\hbar \omega_{P,j}} %
}{\left(1 + r_j\right) \left(1 + {\alpha_j}^2 r_j\right) + e^{\beta\hbar \omega_{P,j}}
},
\label{eq:boltz_nh}
\end{equation}
where $r_j \coloneqq \left(4\omega_{P,j}/\gamma_j\right)^2$.
Here, we eliminated $\gamma_j'$ from the equation in favor of $\alpha_j = \gamma_j/\gamma_j' \in (0, 1)$,
and $\tilde{\beta}$ is the inverse temperature experienced by the system on which the pseudomodes act.

Taking the limit $\alpha_j \to 0$ (equivalently, $\gamma_j' \gg \gamma_j$) in~\eqref{eq:boltz_nh}
recovers the Hermitian single-mode case given in~\eqref{eq:BCF_pseudomode_woRWA} of the main text:
\[e^{\tilde{\beta} \hbar \omega_{P,j}}
    = \frac{\Gamma_j(\omega_{P,j})}{\Gamma_j(-\omega_{P,j})}
    = \frac{(N_j + 1) \mathfrak{L}_j(\omega_{P,j}) + N_j \mathfrak{L}_j(-\omega_{P,j})}{
        (N_j + 1) \mathfrak{L}_j(-\omega_{P,j}) + N_j \mathfrak{L}_j(\omega_{P,j})}
    = e^{\beta \hbar \omega_{P,j}} \frac{(1 + r_j) + e^{-\beta \hbar \omega_{P,j}}}{(1 + r_j) + e^{\beta \hbar \omega_{P,j}}}
.\]
On the other hand, the limit $\alpha_j \to 1$ maximally suppresses the fat tails of the Lorentzians.
The minimum attainable effective temperature in situation can be found by taking
the limit of $\beta \to \infty$ ($T \to 0$) in addition to $\alpha_j \to 1$ in~\eqref{eq:boltz_nh}:
\[T_\text{eff}^\text{(min)}
    = \eval{\frac{\hbar\omega_{P,j}}{\kB \tilde{\beta}}}_{\substack{\alpha_j \to 1\\\beta \to \infty}}
    = \frac{\hbar\omega_{P,j}}{2 \kB \ln(1 + 16 \frac{{\omega_{P,j}}^2}{{\gamma_j}^2})}.\]
Comparing this to the minimum attainable temperature in the corresponding single-pseudomode non-RWA case of the main text,
which may be determined by inserting~\eqref{eq:BCF_pseudomode_woRWA_T0} into~\eqref{eq:T_eff},
\[\frac{\hbar\omega_{P,j}}{\kB \ln(1 + 16 \frac{{\omega_{P,j}}^2}{{\gamma_j}^2})},\]
one finds that the temperature that can be reached by combining the pseudomode with its non-Hermitian counterpart is exactly
half the temperature that can be reached using $\gamma_\text{fixed}$ without non-Hermitian pseudomodes.

The parameters that were used in the example calculation of 5+5 pseudomodes shown in the main text
are given in Tab.~\ref{tab:nh_params}.
\begin{table*}[b]
\centering
\caption{The parameters used in the flat-$T$ 10-pseudomode construction shown in the main text.
\enquote{$g$ scaling factor} refers to the ratio $g_j/g$, where $g_j$ is the coupling strength
of the pseudomode in question and $g$ is the nominal coupling strength used in the main text.
The \enquote{a}~pseudomode is always a standard Hermitian pseudomode while each \enquote{b}~pseudomode
is its corresponding non-Hermitian partner mode.}
\label{tab:nh_params}
\sisetup{round-mode = figures, round-precision = 4}%
\begin{tabular}{c  c c  c c  c c  c c  c c}
\toprule
Pseudomode index & 1a & 1b & 2a & 2b & 3a & 3b & 4a & 4b & 5a & 5b\\
\midrule
$g$ scaling factor              & \num{9.6893714} & \complexnum{4.333218622120165i}
                                 & \num{11.39421111} & \complexnum{10.863954027351495i}
                                  & \num{4.55768444} & \complexnum{4.345581607126747i}
                                   & \num{3.12043323} & \complexnum{2.975216347043787i}
                                    & \num{1.65117806} & \complexnum{1.574336508393114i} \\
$\omega_{P,j}$ in \si{\radian\per\ps}  & \multicolumn{2}{c}{\num{328.50883334}}
                                 & \multicolumn{2}{c}{\num{316.45346331}}
                                  & \multicolumn{2}{c}{\num{263.71121942}}
                                   & \multicolumn{2}{c}{\num{207.2016724}}
                                    & \multicolumn{2}{c}{\num{150.69212538}}\\
$\gamma_j$ in \si{\radian\per\ps}      & \num{45.20763762} & \num{226.03818808}
                                 & \num{113.01909404} & \num{124.32100344}
                                  & \num{180.83055046} & \num{198.91360551}
                                   & \num{169.52864106} & \num{186.48150516}
                                    & \num{237.34009748} & \num{261.07410723}\\
$T$ in \si{\K}                  & \multicolumn{2}{c}{270}
                                 & \multicolumn{2}{c}{345}
                                  & \multicolumn{2}{c}{254}
                                   & \multicolumn{2}{c}{1}
                                    & \multicolumn{2}{c}{250}\\
\bottomrule
\end{tabular}
\end{table*}%

\section{Recipe for constructing the covariance matrix equation} \label{app:recipe}
We state the procedure for converting a standard second-quantization pseudomode master equation
into the matrices that are used as input in sections~\ref{app:second_moments} and~\ref{app:nh_third_quantization}.

Assume our system consists of a single harmonic oscillator (given the index 0)
coupled to $n$ mutually uncoupled pseudomodes (indices 1 through $n$).
Then the Hamiltonian without the RWA is as follows:
\[H = \hbar \omega_0 b^\dag b
    + \sum_{k \geq 1}\left[ \hbar \omega_k {a_k}^\dag a_k
    + \hbar g_k \left(b + b^\dag\right) \left(a_k + {a_k}^\dag\right)\right],\]
where non-real $g_k$ cause the Hamiltonian to become non-Hermitian.
This Hamiltonian corresponds to the following set of matrices:
\[\mathbf{H} = \hbar \begin{pmatrix}
    \omega_0 & g_1 & g_2 & \cdots & g_{n-1} & g_n\\
    g_1 & \omega_1 & 0 & \cdots & 0 & 0\\
    \vdots  & & \ddots & & \vdots\\
    g_{n-1} & 0 & 0 & \cdots & \omega_{n-1} & 0\\
    g_{n}   & 0 & 0 & \cdots & 0 &\omega_{n}
\end{pmatrix},
\qquad
\mathbf{K} = \frac{\hbar}{2}\begin{pmatrix}
    0 & g_1 & g_2 & \cdots & g_{n-1} & g_n\\
    g_1 & 0 & 0 & \cdots & 0 & 0\\
    \vdots  & & \ddots & & \vdots\\
    g_{n-1} & 0 & 0 & \cdots & 0 & 0\\
    g_{n}   & 0 & 0 & \cdots & 0 & 0
\end{pmatrix}
    = \overline{\mathbf{J}}.\]
The above matrices do not include the Lamb-shift-like counterterms discussed in sec.~\ref{app:counterterms}.
We can include these in the following sense,
\begin{multline*}
\sum_k \left( \frac{{p_k}^2}{2m_k} + \frac{m_k {\omega_k}^2}{2} {x_k}^2\right)
    + \sum_{\substack{k, l\\ l > k}} \frac{\hbar g_{k,l}}{2 \lambda_k \lambda_l} \left(x_k - x_l\right)^2\\
    =
\sum_k \hbar \omega_k \left({a_k}^\dag a_k + \frac{1}{2}\right)
    + \sum_{\substack{k, l\\ l \neq k}} \frac{\hbar g_{k,l}}{2} \sqrt{\frac{m_l \omega_l}{m_k \omega_k}} \left({a_k}^2 + \left({a_k}^\dag\right)^2 + 2 {a_k}^\dag a_k + 1\right)
        - \sum_{\substack{k, l\\ l > k}} \hbar g_{k,l} \left(a_k + {a_k}^\dag\right) \left(a_l + {a_l}^\dag\right)\\
    =
    \sum_k \hbar \underbrace{\sqrt{{\omega_k}^2 + \sum_{l \neq k} \frac{\hbar g_{k,l}}{m_k \lambda_k \lambda_l}}}_{\eqqcolon \omega_k'}
                \left({a_k}^\dag a_k + \frac{1}{2}\right)
        - \sum_{\substack{k, l\\ l > k}} \hbar g_{k,l} \left(a_k + {a_k}^\dag\right) \left(a_l + {a_l}^\dag\right)
\end{multline*}
by modifying the diagonal elements in $\mathbf{H}$ to use the renormalized frequencies $\omega_k'$
instead of the bare frequencies $\omega_k$.
Note that these frequencies can be expanded to
\[\omega_k' = \sqrt{{\omega_k}^2 + 2 \sum_{l \neq k} g_{k,l} \sqrt{\frac{m_l}{m_k} \omega_k \omega_l}}.\]
For the specific set of couplings $g_{k,l}$ that we have in our model, this results in:
\[\omega_0' = \sqrt{{\omega_0}^2 + 2 \sum_{1 \leq l \leq n} g_l \sqrt{\frac{m_l}{m_0} \omega_0 \omega_l}}
\qand
\omega_k' = \sqrt{{\omega_k}^2 + 2 g_k \sqrt{\frac{m_0}{m_k} \omega_k \omega_0}} \qfor{k > 0}. \]
We will assume that all masses $m_k$ are identical.
Note, however, that including imaginary-valued $g_l$ in such a procedure will cause the local
frequencies $\omega_k'$ to also become imaginary, which is likely not a desirable property.

The remaining matrices $\mathbf{M}$, $\mathbf{N}$, $\mathbf{L}$ are given by the Lindblad dissipators.
For non-zero temperatures, we have the following $2n$ separate jump operators with $j \in \{1, 2, \ldots n\}$:
\[\underline{l}_{2j} = \sqrt{\frac{\gamma_j}{2} (N_j + 1)} \underline{e}_j,
    \quad \underline{k}_{2j}  = 0,
    \qand \underline{l}_{2j + 1}  = 0,
    \quad \underline{k}_{2j + 1} = \sqrt{\frac{\gamma_j}{2} N_j} \underline{e}_j, \]
where $\underline{e}_j$ is the unit vector that is 1 at position $j$.
This results in
\begin{align*}
\mathbf{M} &= \sum_\mu \underline{l}_{\mu} \otimes \overline{\underline{l}}_{\mu}
    = \frac{1}{2} \diag(0, \gamma_1 (N_1 + 1), \gamma_2 (N_2 + 1), \ldots, \gamma_n (N_n + 1)),\\
\mathbf{N} &= \sum_\mu \underline{k}_{\mu} \otimes \overline{\underline{k}}_{\mu}
    = \frac{1}{2} \diag(0, \gamma_1 N_1, \gamma_2 N_2, \ldots, \gamma_n N_n),\\
\mathbf{L} &= \sum_\mu \underline{l}_{\mu} \otimes \overline{\underline{k}}_{\mu}
    = 0.
\end{align*}
We then insert these matrices into~\eqref{eq:Xmat} and \eqref{eq:Ymat} and solve Eq.~(23) of Ref.~\cite{Prosen2010}, i.e.,
\[\mathbf{X}^\text{T} \mathbf{Z} + \mathbf{Z} \mathbf{X} = \mathbf{Y},\]
for $\mathbf{Z}$ to obtain the results as shown in~\ref{app:second_moments}.
If we are only interested in the properties of the oscillator with index~0,
then we can extract the relevant correlation functions from the matrix $Z$,
\[\expval{b^2} = Z_{0,0},
\quad \expval{\left(b^\dag\right)^2} = Z_{n+1,n+1} = {Z_{0,0}}^\ast,
    \qand \expval{b^\dag b} = Z_{0, n+1} = Z_{n+1, 0}.\]

\end{document}